\documentclass[
twocolappendix, twocolumn, floatfix]{aastex631}
\usepackage{amsmath}

\usepackage{comment}

\begin{document}

\title{Microlensing at Cosmological Distances: Event Rate Predictions in the Warhol Arc of MACS 0416}

\correspondingauthor{J.M. Palencia}
\author[0000-0003-0942-817X]{J.M. Palencia}
\affiliation{IFCA, Instituto de F\'isica de Cantabria (UC-CSIC), Av. de Los Castros s/n, 39005 Santander, Spain}
\email{palencia@ifca.unican.es}

\author[0000-0001-9065-3926]{J.M. Diego}
\affiliation{IFCA, Instituto de F\'isica de Cantabria (UC-CSIC), Av. de Los Castros s/n, 39005 Santander, Spain}

\author[0000-0003-2091-8946]{L. Dai}
\affiliation{Department of Physics, University of California, 366 Physics North MC 7300, Berkeley, CA 94720, USA}

\author[0000-0002-2282-8795]{M. Pascale}
\affiliation{Department of Astronomy, University of California, 501 Campbell Hall \#3411, Berkeley, CA 94720, USA}

\author[0000-0001-8156-6281]{R. Windhorst}
\affiliation{School of Earth and Space Exploration, Arizona State University, Tempe, AZ 85287-1404, USA}

\author[0000-0002-6610-2048]{A. M. Koekemoer}
\affiliation{Space Telescope Science Institute, 3700 San Martin Drive,
Baltimore, MD 21218, USA}

\author[0000-0002-4490-7304]{Sung Kei Li}
\affiliation{Department of Physics, The University of Hong Kong, Pokfulam Road, Hong Kong}

\author[0000-0002-3634-4679]{B. J. Kavanagh}
\affiliation{IFCA, Instituto de F\'isica de Cantabria (UC-CSIC), Av. de Los Castros s/n, 39005 Santander, Spain}

\author[0000-0002-4622-6617]{Fengwu Sun}
\affiliation{Center for Astrophysics Harvard \& Smithsonian: Cambridge, Massachusetts, US}


\author[0000-0003-1276-1248]{Amruth Alfred}
\affiliation{Department of Physics, The University of Hong Kong, Pokfulam Road, Hong Kong}

\author[0000-0002-7876-4321]{Ashish K. Meena}
\affiliation{Department of Physics, Ben-Gurion University of the Negev, PO Box 653, Be’er-Sheva 8410501, Israel}

\author[0000-0002-8785-8979]{Thomas J. Broadhurst}
\affiliation{Department of Theoretical Physics, University of Basque Country UPV/EHU, Bilbao, Spain}
\affiliation{Ikerbasque, Basque Foundation for Science, Bilbao, Spain}
\affiliation{Donostia International Physics Center, Paseo Manuel de Lardizabal, 4, San Sebasti\'an, 20018, Spain}

\author[0000-0003-3142-997X]{Patrick L. Kelly}
\affiliation{Minnesota Institute for Astrophysics, 
University of Minnesota, 116 Church St. SE, Minneapolis, MN 55455, USA}

\author[0000-0002-4693-0700]{Derek Perera}
\affiliation{Minnesota Institute for Astrophysics, 
University of Minnesota, 116 Church St. SE, Minneapolis, MN 55455, USA}

\author[0000-0002-1681-0767]{Hayley Williams}
\affiliation{Minnesota Institute for Astrophysics, 
University of Minnesota, 116 Church St. SE, Minneapolis, MN 55455, USA}

\author[0000-0002-0350-4488]{Adi Zitrin}
\affiliation{Department of Physics, Ben-Gurion University of the Negev, PO Box 653, Be’er-Sheva 8410501, Israel}


\begin{abstract}

Highly magnified stars ($\mu$ $>$ 100) are now routinely identified as transient events at cosmological distances thanks to microlensing by intra-cluster stars near the critical curves of galaxy clusters.
Using the {\it James Webb} Space Telescope (JWST) in combination with the {\it Hubble} Space Telescope (HST), we outline here an analytical framework that is applied to the Warhol arc (at $z=0.94$) in the MACS 0416 galaxy cluster (at $z=0.396)$ where over a dozen microlensed stars have been detected to date. This method is general and can be applied to other lensed arcs.
Within this lensed galaxy we fit the spatially resolved SED spanned by eight JWST-NIRCam filters combined with three ACS filters, for accurate lensed star predictions in 2D.
With this tool we can generate 2D maps of microlensed stars for well resolved arcs in general, incorporating wavelength dependence and limiting apparent magnitude. These maps can be directly compared with planned cadenced campaigns from JWST and Hubble, offering a means to constrain the IMF and the level of dark matter substructure.

\end{abstract}


\section{Introduction} \label{sec:intro}

Galaxy clusters are the most powerful gravitational lenses in the Universe.
Comprising hundreds to thousands of galaxies, hot gas, and dark matter, these clusters distort space-time, bending the paths and amplifying the  light of distant galaxies behind them.
This well-known gravitational lensing effect results in a range of phenomena for background sources, such as altering their apparent position and shape, amplifying their brightness, and creating multiple images with different arrival times~\citep{Schneider1992}.

Galaxies near the critical curves (hereafter CCs) of galaxy clusters form elongated images, commonly known as lensed arcs, which are typically magnified by a factor of $\mathcal{O}(10)$. CCs map onto caustics in the source plane with a well defined dependency of the magnification, $\mu$, with respect to the distance, $d$, to the caustic, $\mu=\mu_o/\sqrt{d}$, where for clusters, $\mu_o$ usually takes values $\sim$ 10--30 when $d$ is expressed in arcseconds. 
Sources with radius $R$ (such as compact star forming regions where $R\lesssim 0.1"$) within a lensed galaxy crossing a  caustic, can achieve maximum magnifications $\mu_{\rm max}=\mu_o/\sqrt{R}\approx 50$. 
Even smaller objects, such as stars  $R\lesssim 10^{-8}$ arscec), can theoretically achieve extreme magnification factors  $\mu_{\rm max}\sim10^6$ \citep{Miralda-Escude1991}. Such large magnification is possible if the distribution of mass is smooth and the lensed image lies extremely close to the CC. 
However, microlenses (such as stars from the intracluster medium) and small-scale structures within the cluster (dwarf galaxies, globular clusters, subhalos predicted in abundance by massive particle CDM, or pervasive density modulations predicted in ultralight particle CDM or Wave DM) perturb the gravitational potential, reducing this maximum achievable magnification to the order of tens of thousands \citep{Venumadhav2017,Diego2018,Weisenbach2024}. This reduction in magnification near the CC is compensated (as demanded by flux conservation) by larger magnification factors around the microlenses and small-scale substructures farther away from the CC. These small scale structures create a web of micro- and millicaustics near the cluster caustic. 

When a background star moves through this web of microcaustics the combined lensing effect from different microlenses produces many unresolved images (microimages) of the same star, resulting in a combined magnification typically in the order of thousands, hence the term ``extremely magnified stars''.
The angular separation between these images is typically on the scale of microarcseconds~\citep{Venumadhav2017}, hence far smaller than the resolution capabilities of current telescopes that operate with resolutions of tens of milliarcseconds for the most powerful optical and infrared telescopes. As a result, we observe a single image of the background star, which represents the combined magnification (or flux) of all the microimages.
Due to the relative motion between the source and the web of microcaustics, this magnification varies over time, peaking when the source crosses one of the multiple microcaustics and then gradually decreasing until the star is no longer detectable.
These events can last from a few days to several years, depending primarily on the redshift of the source, its luminosity, the mass of the microlens, and more importantly the relative velocity and direction of motion with respect to the web of microcaustics. 

Microlenses can also be made of compact dark matter objects, such as Primordial Black Holes (PBHs) \citep{Venumadhav2017,Diego2018,Carr2020,Green2021,Palencia2024}.
However, current studies are consistent with a fully stellar microlens scenario \citep{Oguri2018,Muller2025}.
The discovery of Icarus \citep{Kelly2018}, a lensed star at redshift $z=1.49$, marked the beginning of this field.
Since then, $O(100)$  of these lensed high-redshift stars have been discovered with both the {\it Hubble} Space Telescope (HST) and the {\it James Webb} Space Telescope (JWST).

Notable discoveries include Earendel, the farthest known star with a redshift of $z\approx6$ \citep{Welch2022a, Welch2022b} (see \citet{Ji2025} for caveat on the source size constraint), and the Dragon Arc in Abell 370, with 46 detected events in a single galaxy \citep{Fudamoto2024}.
These discoveries open a new frontier in gravitational lensing, allowing us to study stars in the early stages of the Universe post-reionization, their evolution, the possible first generation of Population III stars \citep{Zackrisson2024}. In addition the nature of dark matter, particularly its presence in the form of compact objects within galaxy clusters \citep{Diego2018,Oguri2018} or ubiquitous fluctuations in the DM Density as predicted by wave DM models \citep{Amruth2023, Broadhurst2025}. High-cadence light curves of caustic crossing events can serve as a probe for DM micro-structures formed in the early universe~\citep{Dai2020a}.

MACS J0416.1-2403 \citep{Ebeling2001, Mann2012}, commonly referred to as MACS 0416, at redshift $z=0.396$, is one of six galaxy clusters studied by HST as part of the {\it Hubble} Frontier Fields program \citep[HFF,][]{Lotz2017}.
Later observed by JWST, MACS 0416 holds the record for the lens with the largest number of spectroscopically confirmed lensed galaxies \cite{CANUCS2024}. Several of the lensed galaxies cross the CC, hence with portions of these galaxies reaching maximum magnification.

This cluster has been extensively studied in the optical and infrared (IR) bands, with data from both telescopes being used by experts to derive best-fit lens models.
Prior to JWST observations of this cluster, some of the earliest lensed stars at cosmological distances ($z \gtrsim 1$) were discovered in this cluster by HST~\citep{Chen2019,Kaurov2019}. Later on, Flashlights, an HST program for taking extremely deep images of lensing galaxy clusters, identified about a dozen additional lensed stars~\citep{Kelly2022} in two background galaxies, named Spock and Warhol, both at redshift $z \approx 1$.
Future deeper observations of these galaxies are expected to yield yet more microlensing events.

The modest redshift of $z \approx 1$ for Spock and Warhol, with a distance modulus of 44.10, increases the detection rates, since we can observe fainter stars.
Typically, observations are biased towards the brightest stars, as they require a smaller (and more likely) magnification, in order to be detectable.
This makes these two galaxies prime targets for studying the abundance of different star types at redshift $z=1$, specifically probing the high-mass end of the Initial Mass Function (IMF), as we are limited to the most luminous and hence brightest stars.

In this paper, we focus on Warhol and provide a detailed forecast for the number of detectable events across eight NIRCam filters.
We also present the first spatial distribution forecast for the number of events.
Previous studies have made similar predictions for Spock \citep{Diego2024a}.
Warhol and Spock are not the only galaxies in MACS 0416 that host microlensing events.
Mothra \citep{Diego2023b}, a lensed star at redshift $z=2.09$, was also discovered in a galaxy behind MACS 0416, providing interesting constraints on dark matter, as it demands the presence of a small millilens near the lensed star in order explain its peculiar magnification.

Future dedicated observations of this galaxy cluster are expected to lead to many more detections, offering valuable insights into the nature of dark matter within galaxy clusters, as well as the formation and evolution of stars at high redshift. A larger number of events will allow to test the results derived from this paper.

This paper is structured as follows. Section~(\ref{sec:data}) details the data used in our analysis.
In Section~(\ref{sec:methodology}), we outline the methodology used to transition from data analysis to forecasting the rate of lensed events.
The results, covering both spatial and integrated event counts across eight NIRCam filters, are presented in Section~(\ref{sec:results}).
In Section~(\ref{sec:discussion}), we discuss the prospects for detecting lensed high redshift stars in MACS 0416 and similar arcs.
Finally, our conclusions are summarised in Section~(\ref{sec:conclusion}).

Unless stated otherwise, we assume a flat cosmological model with $\Omega_m=0.3$, $\Lambda=0.7$, and $h=0.7$ (100 km s$^{-1}$ Mpc$^{-1}$). All magnitudes are given in the AB system. The Warhol arc is at redshift $z=0.94$, within the adopted cosmological model 1 arcsec corresponds to~$\sim15.3$ kpc.

\section{Data} \label{sec:data}
In this section, we describe the high-resolution {\it HST} and {\it JWST} mosaics of MACS 0416 which we use, also the scientific programs from which the data were obtained, and present the lens model used in this work.

To perform all photometric SED fitting, we used a combination of 11 photometric mosaic images of MACS 0416, using public data from {\it HST} (described further below) and the {\it JWST} Prime Extragalactic Areas for Reionization and Lensing Science (PEARLS) program \citep{Windhorst2023}. We focus on a 24-arcsecond region centred on Warhol, using mosaics at a pixel scale of 30 mas per pixel. 

For the lensing model of MACS 0416, we used the free-form lens model from \cite{Diego2024b} based on the latest {\it JWST} data.  
For the contribution of microlenses at the position of Warhol we assumed a mean surface mass density of microlenses, $\Sigma_{\ast} = 59.39\, {\rm M}_\odot\, {\rm pc}^{-2}$~\citep{Kaurov2019}, which is subsequently modulated by the intra-cluster light (ICL). The rapidly declining brightness near the BCG suggests that the surface mass density varies along the arc, rather than remaining constant.

\subsection{\it HST}
We use {\it HST} mosaics that were constructed by PEARLS team members following the techniques first described by \citet{Koekemoer2011,Koekemoer2013}, where these mosaics include public archival {\it HST} data on MACS 0416, specifically data from the Hubble Frontier Fields \citep[HFF,][]{Lotz2017}, also the Beyond Ultradeep Frontier Fields and Legacy Observations program \citep[BUFFALO,][]{Steinhardt2020}, and other public archival {\it HST}  programs available in the STScI MAST Archive\footnote{\url{https://archive.stsci.edu}}. For this paper, the mosaics that we use include all the {\it HST} ACS observations on this cluster in the F435W, F606W, and F814W filters, covering a wavelength range of approximately 0.35 to 0.96 microns. 

\begin{figure}[ht!]
    \centering
    \includegraphics[width = \linewidth]{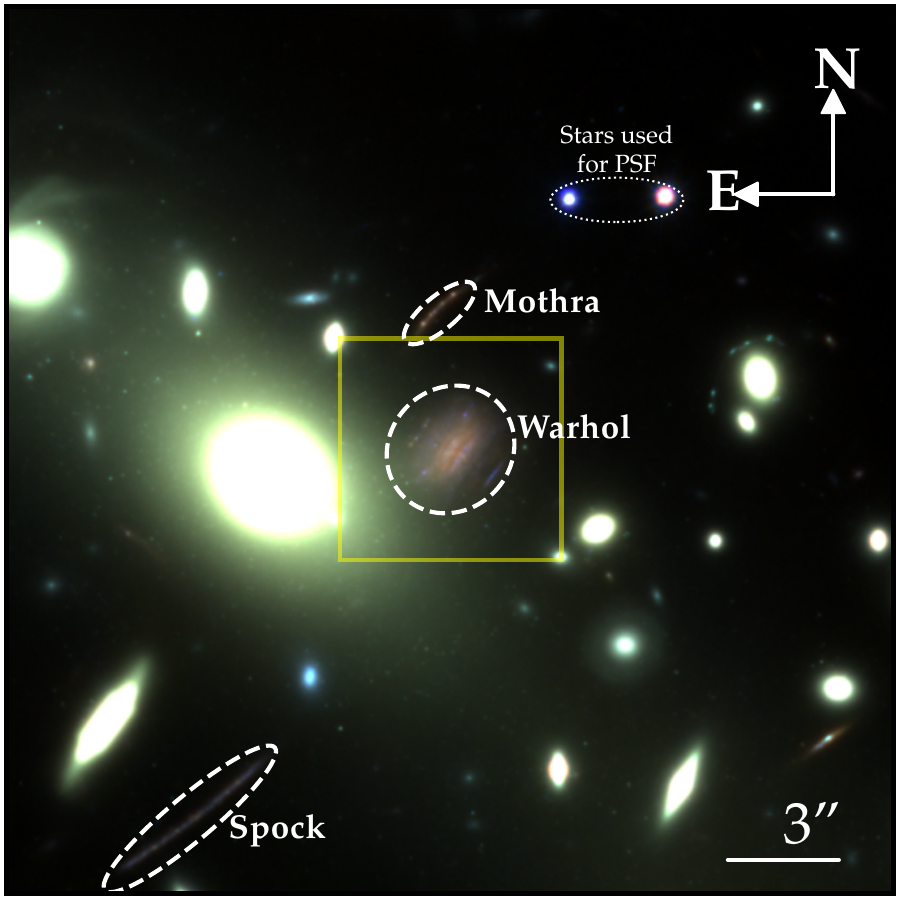}
    \caption{
    MACS 0416 colour image showing a 24"$\times$ 24" region centered on the Warhol arc. Other prominent caustic crossing galaxies where microlensing transients have been identified are also marked.
    The colour image was created by combining filters from two instruments: NIRCam (F444W and F356W for the red channel; F277W, F200W, and F150W for the green channel; F115W and F090W for the blue channel) and WFC3 (F814W, F606W, and F435W for the blue channel). The yellow square outlines the central region analysed in this study.
    }
    \label{fig: MACS 0416}
\end{figure}

\subsection{\it JWST}

We use {\it JWST} mosaics constructed by PEARLS team members, using data from the PEARLS program obtained in eight filters from the NIRCam instrument: F090W, F115W, F150W, F200W, F277W, F356W, F410M, and F444W, covering wavelengths from 0.795 to 4.981 microns.
The corresponding magnitude depths, at which a point-source signal-to-noise ratio of 5 is achieved for fake sources injected~\citet{Williams+inprep}, are 29.7, 29.5, 29.5, 29.5, 29.5, 29.6, 29.0, and 29.3 $\,m_{\textrm{AB}}$, respectively. The {\it HST} and {\it JWST} mosaics that we use were all produced in a way that aligns them to a common astrometric grid, at a pixel scale of 30 mas per pixel. See \citet{Windhorst2023} for further details about the construction of the mosaics that we use here.

\subsection{\it Lens model}
We utilise the free-form code WSLAP+~\citep{Diego2005,Diego2007} to derive the mass distribution, and magnification of the macromodel. WSLAP+ is a hybrid  method that allows to combine weak and strong lensing constraints. For MACS0416, only strong lensing constraints are used. The method exploits the linearity (with mass) of the deflection field, which can be expressed as a linear combination of functions of the mass. It describes the mass distribution as the combination of two components: (i) a superposition of Gaussians located either on a regular grid or a adaptive grid, with additional Gaussians or increased resolution in specific regions of the cluster where higher mass is located, effectively acting as a prior for the mass distribution. This represents the smooth global component of the cluster lens. (ii) A compact component that accounts for the individual masses of cluster members, assumed to be proportional to their flux in a certain reference band. For this component, there is only one free parameter: the mass scaling factor or mass-to-light ratio of the cluster members. Different layers can be introduced to account for variations in the mass-to-light ratio among different types of galaxies within the cluster.
The strong- and weak-lensing problem is then formulated as a system of linear equations:
\begin{equation}
    \bf{\Theta = \Gamma X},
\end{equation}
where the array $\bf{X}$ contains the free parameters: the grid points for the Gaussian approximation of the smooth component, the mass-to-light scalings, and the identified sources positions. The lensing observables, including the positions of strongly lensed sources (and when available the two shear components), are arranged in the $\bf{\Theta}$ array. The $\bf{\Gamma}$ matrix is known and depends on the fiducial grid and redshifts of the lensed galaxies. The solution $\bf{X}$ is obtained by inverting a scalar function constructed from the system of linear equations. The lens model for MACS0416 is derived from a very large number of constraints (343) and is described in detail in \cite{Diego2024b}.

\section{Methodology} \label{sec:methodology}
The objective of this paper is to derive a realistic forecast for the expected number of microlensing events on the Warhol arc. These transient events originate from stars within the lensed galaxy whose apparent magnitude, $m$, is typically well below the detection threshold, making them undetectable in most observational epochs. However, a temporary magnification boost, $\mu_{\rm micro}$, arises when the source approaches a micro caustic, which brings their apparent magnitude, $m_{\rm obs}$, above the detection threshold:
\begin{equation}
m_{\rm obs} = m - 2.5\textrm{log}_{10}(\mu_{\rm micro}) \lesssim m_{\varepsilon},
\end{equation}
where $m_{\varepsilon}$ represents the detection threshold. The average number of detected stellar transient events depends on two main factors: the micro-magnification probability distribution function (PDF) and the background stellar population. The former describes the micro-magnification as a random variable that depends on the tangential component of the macrolens, $\mu_{\rm t}$, the radial component, $\mu_{\rm r}$, and the surface mass density of microlenses, $\Sigma_{\ast}$ \citep[see][for further details]{Palencia2024}. Furthermore, local perturbations in the macrolens caused by millilenses can reshape the local macro-magnification map, thereby altering the magnification PDFs. The latter is influenced by various parameters, such as the age of the population, its metallicity, environmental constraints like dust attenuation, and the total stellar mass, among others. These parameters collectively shape the SED of the population, though some degeneracies exist, such as the dust-SFH degeneracy where dusty and older populations both produce similar features with attenuated UV-rest frame emission. Consequently, the number and properties of the source stars can be inferred from the total SED measured from the galaxy.

In this section, we present our methodology: beginning with a Markov Chain Monte Carlo (MCMC) parameter estimation of the stellar population parameters derived from SEDs across different regions of Warhol, which allows us to construct a set of source stellar populations. We then integrate local magnification PDFs that reflect the peculiarities of the lens model, applying them to the source stars to forecast the average expected number of events in various NIRCam filters at multiple detection thresholds. The results are presented and further discussed in sections~(\ref{sec:results}) and~(\ref{sec:discussion}), respectively.

\subsection{\it SED Fitting}
\begin{figure*}[ht!]
    \centering
    \includegraphics[width = \linewidth]{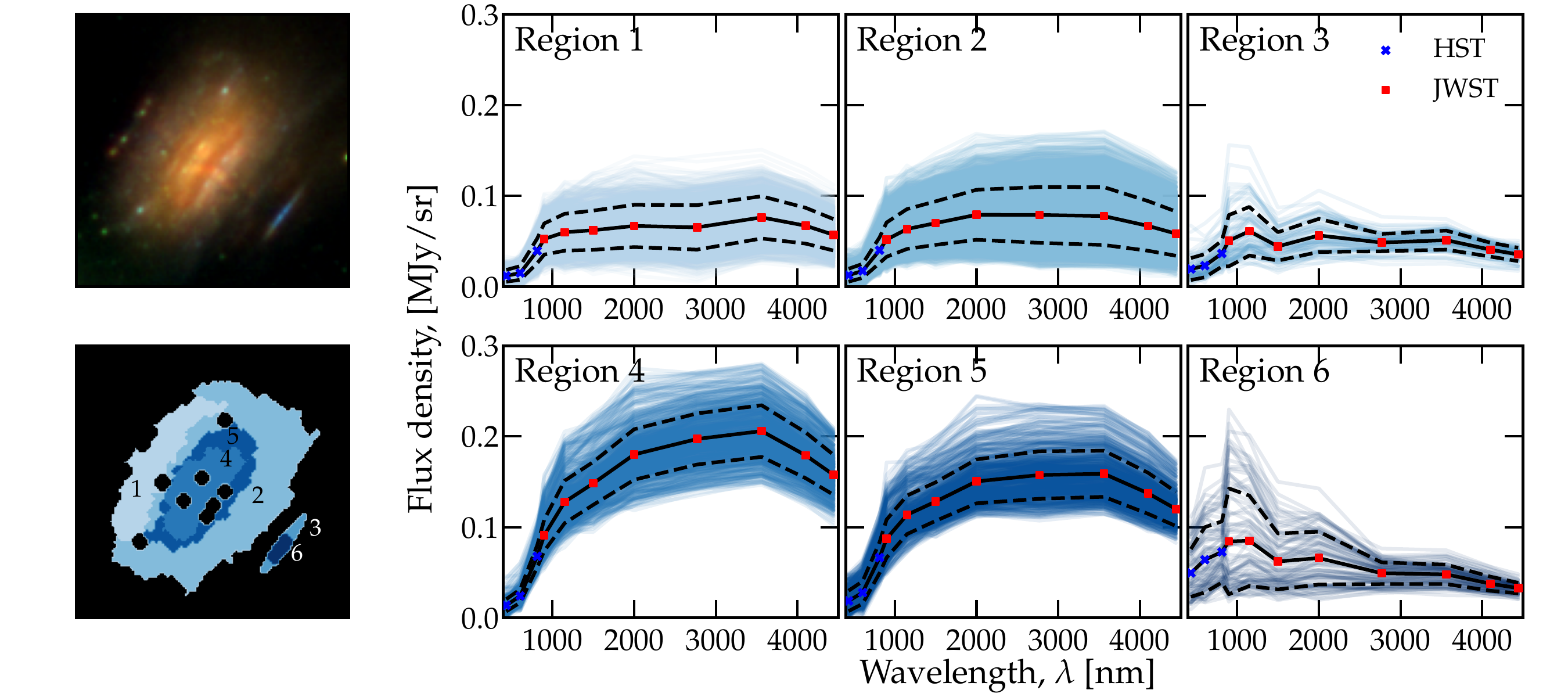}
    \caption{
    Similar SED regions after clustering. {\em Top-left panel: \/} Colour image of Warhol. {\em Bottom-left panel: \/} Regions of similar SED pixels on Warhol. Black dots indicate masked regions, globular clusters from the intracluster medium, or transient events whose SEDs do not represent the underlying stellar population. {\em Second, third, and fourth columns: \/} SEDs for each pixel in the different regions. The solid black line represents the median SED, while the dashed black lines indicate the 1$\sigma$ contour. Violet crosses and red squares indicate the photometric measurements obtained using HST and JWST filters, respectively, all introduced in Section~\ref{sec:data}.
    }
    \label{fig: Regions}
\end{figure*}
The first step in our analysis involves retrieving the SED of the Warhol arc. To achieve this, we carefully remove the contaminating foreground light from the galaxy cluster, specifically the ICL. We utilise {\tt SExtractor}~\citep{Bertin1996}, a publicly available code that enables source detection, deblending of overlapping or nearby sources, and image background estimation. Additionally, {\tt SExtractor} was used to delineate the region of pixels corresponding to the Warhol arc. The next step is to subtract the remaining part of the brightest cluster galaxy (BCG) emission, which {\tt SExtractor} identifies as a source and treats independently from the image background. In our specific case, the BCG is almost directly in front of Warhol, so its subtraction is crucial for recovering an accurate representation of Warhol's SED. For this task, we rely on {\tt PetroFit}~\citep{Geda2022}, a Python package designed for tasks such as galaxy light profile fitting. We modelled the main BCG using a Sérsic profile~\citep{Sersic1963}, which was then subtracted from the images. This removal of both foreground contaminants was performed separately for each filter image, resulting in a set of contamination-free images across different filters for the Warhol arc.

The next step is to match the Point Spread Function (PSF) of our images to prevent biases and ensure consistent flux measurements at the pixel level, thereby guaranteeing accurate photometry. Since broader PSFs spread the light of sources more, matching the images to the broadest PSF among all the filters ensures that we achieve bias-free photometry in each pixel of the arc. To accomplish this, we first need to determine the PSF for each filter. Foreground stars, being point-like sources (significantly smaller than the instrument's response), directly reflect the shape of the instrument's PSF in their images. We use two stars located north-west of Warhol (see Figure~\ref{fig: MACS 0416}) to construct an empirical PSF using {\tt Photutils}~\citep{Bradley2024}. Next, we calculate a convolution kernel for each image. This kernel is the inverse Fourier transform of the ratio between the Fourier transforms of the PSFs, specifically the narrower PSF over the broader one. We then convolve the images with narrower PSFs using the corresponding kernels, resulting in a set of PSF-matched, contaminant-free images that are ready for accurate photometric estimation.

The most accurate estimation of the background stellar population would involve fitting the SED at each pixel independently. However, as our method relies on MCMC to find the set of stellar population parameters that best reproduce the SEDs, it would be extremely computationally demanding. Instead, we group pixels with similar SEDs. To refine this process, we first masked the known brightest globular clusters (GCs) and potential transient events in our images, these appear as black dots in Figure~\ref{fig: Regions}.

GCs in MACS 0416 present challenges for observing microlensing transients behind them, but their local modifications to the macromodel and enhanced microlens density can significantly affect the rate of events around such millilenses.
As for pixels containing the brightest transients in Warhol, we mask them because their SEDs do not represent the underlying stellar population, the transient's flux outshines the background, preventing accurate SED determination. Both GCs and bright transients appear as local maxima, making their identification straightforward. We remove an area equivalent to twice the $\sigma$ of an effective Gaussian centred at each maximum. Since the masked areas are relatively small compared to the size of the arc, we do not anticipate a significant underestimation of the total number of transient events expected in the arc, though recovering the full spatial distribution of events within the arc remains challenging.

\begin{figure*}[ht!]
    \centering
    \includegraphics[width = \linewidth]{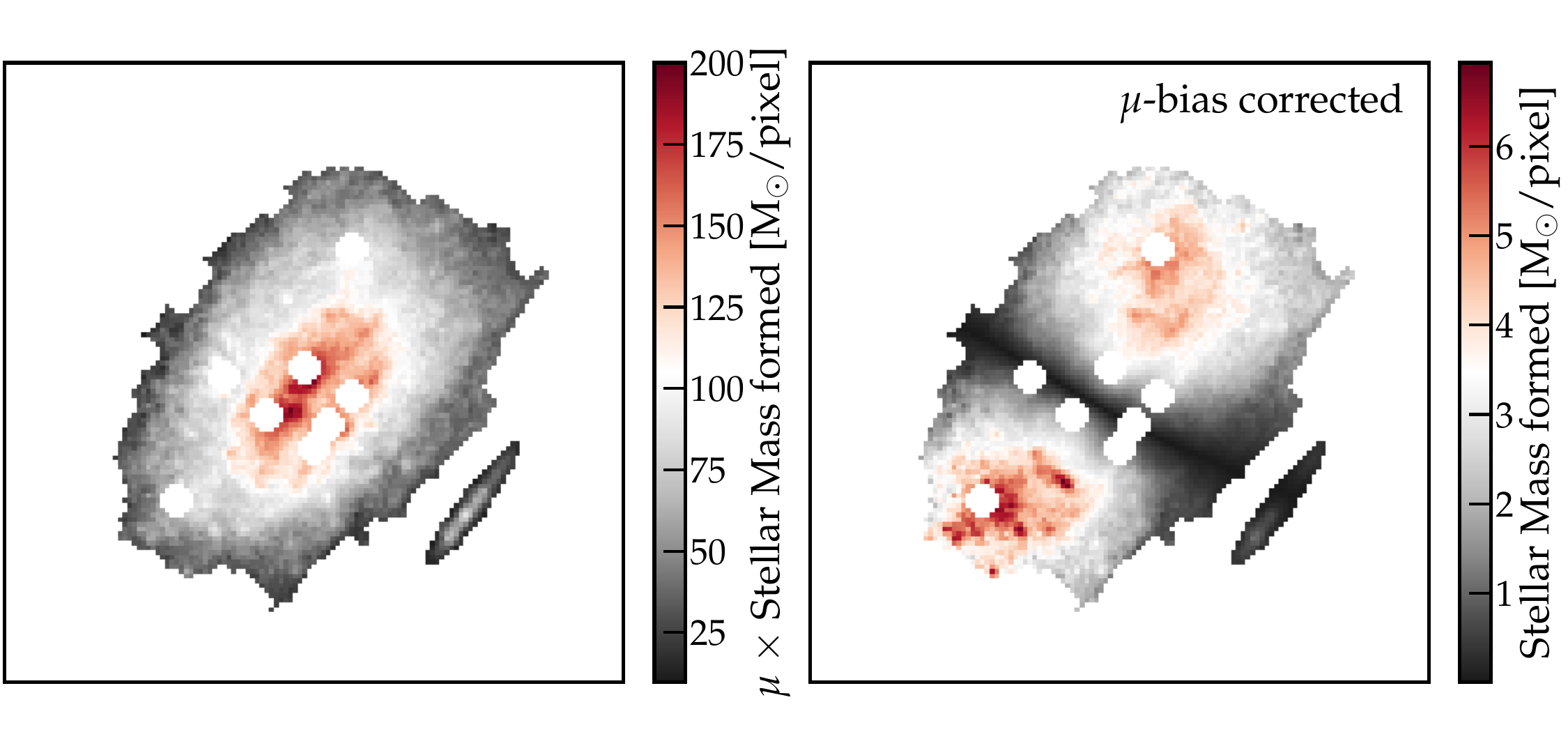}
    \caption{
    Distribution of total stellar mass formed per pixel in Warhol throughout the entire SFH shown before and after accounting for magnification bias correction, in the left and right panels respectively. Brighter pixels may suggest a larger stellar mass, but magnification can make smaller stellar populations appear brighter, introducing a bias. To correct for this effect, we divide the stellar mass formed in each pixel by $\mu_{\rm m}$, which accounts for the change in source-plane area relative to the lens-plane pixel area due to magnification. Near the critical curve, the number of stars in the source plane imaged in this region is lower than in areas with lower magnification, but the magnification is extremely large, enabling the detection of individual fainter stars but within a reduced source area. One pixel corresponds to 32 mas, or 490 pc on the source plane if $\mu=1$. 
    }
    \label{fig: mass_over_mu}
\end{figure*}

Once transients and GCs are masked, we apply various methods to group pixels with similar SEDs. First, we perform a Principal Component Analysis (PCA) to reduce the dimensionality of our dataset from 11 filters to 3 principal components, which capture 96.6\% of the colour variability in our data. Additionally, we conduct a colour-colour analysis by plotting the differences between pairs of filters. Both PCA and colour-colour analysis can be combined with clustering techniques such as K-means, DBSCAN, or by applying linear cuts. We cluster the pixels based on two criteria: cuts performed in the reduced data spaces from PCA and colour-colour analysis, and spatial correlation to ensure that regions are spatially connected. 
The final regions identified through clustering, along with their median SEDs and 1$\sigma$ contours, are shown in Figure~\ref{fig: Regions}.

To estimate the photometric error, we use aperture photometry with different apertures around the Warhol arc. By analysing the observed variability in flux/area measurements around the arc, we can infer an approximate photometric error per square arcsecond. This error can then be extrapolated to the Warhol regions with known areas.

With accurate photometric SEDs for specific regions, calculated as the total contribution from all pixels within that region, and estimates of the flux errors for each filter, we can now fit these SEDs to reference models. For this purpose, we use the Flexible Stellar Population Synthesis ({\tt FSPS}) library~\citep{Conroy2009, Conroy2010a, Conroy2010b, pythonFSPS}. {\tt FSPS} is a sophisticated tool for modelling the SED of stellar populations, offering flexible input options that allow users to specify various parameters, including IMFs, metallicities, masses, redshifts, ages, and more. It also allows for the incorporation of environmental conditions, such as dust attenuation, dust and nebular emission, or even emission from active galactic nuclei (AGN) with different torus optical depths, which shape the AGN's contribution to the total SED.

One of the many outputs that {\tt FSPS} generates is the photometric fluxes measured in the filters we are using. This enables us to compare model predictions to our data and explore the parameter space to find the best-fit parameters that produce an SED that minimises a $\chi^2$ with our observations. We perform this fitting for all our regions, assuming different Star Formation History (SFH) models, which will influence the quantity and types of stars in our selected regions.

The fitting process is carried out using modified packages from the {\tt PixedFit}\citep{Abdurro'uf2021} library, which utilises {\tt FSPS} to generate the models. Additionally, we employ {\tt emcee} \citep{Foreman-Mackey2013}, a Python implementation of the Affine Invariant MCMC Ensemble sampler~\citep{Goodman2010}, to explore the parameter space and find the best-fit solutions.

Once the MCMC fitting is completed, we reconstruct the SFH of the stellar population based on the assumed model and best-fit parameters. This provides the star formation rate, i.e., the total amount of stellar mass formed per year from the beginning of the SFH to the current age of the galaxy at its redshift, $z \sim 1$. By integrating the SFH, we can determine the total amount of stars within various age bins. Since we used a region-integrated SED and assumed a homogeneous population of stars throughout the region, the mass distribution across the region is derived by weighting the total mass per age bin for that specific region by the flux distribution.

Once this process is completed for all six regions, we can produce a map containing the mass distribution for each age bin (see figure~\ref{fig: mass_over_mu} for the integrated value over all SFH). As the total mass distribution is weighted by the flux distribution, the mass obtained is biased by lensing magnification. To correct for this, we divide the mass in each pixel by its magnification. Since higher magnification corresponds to a smaller area in the source plane, this correction naturally reduces the mass in regions of high magnification. Conversely, higher magnifications make it easier to detect events in these regions. The corrected version of the mass distribution is shown in figure~(\ref{fig: mass_over_mu}).

\begin{figure}[ht!]
    \centering
    \includegraphics[width = \linewidth]{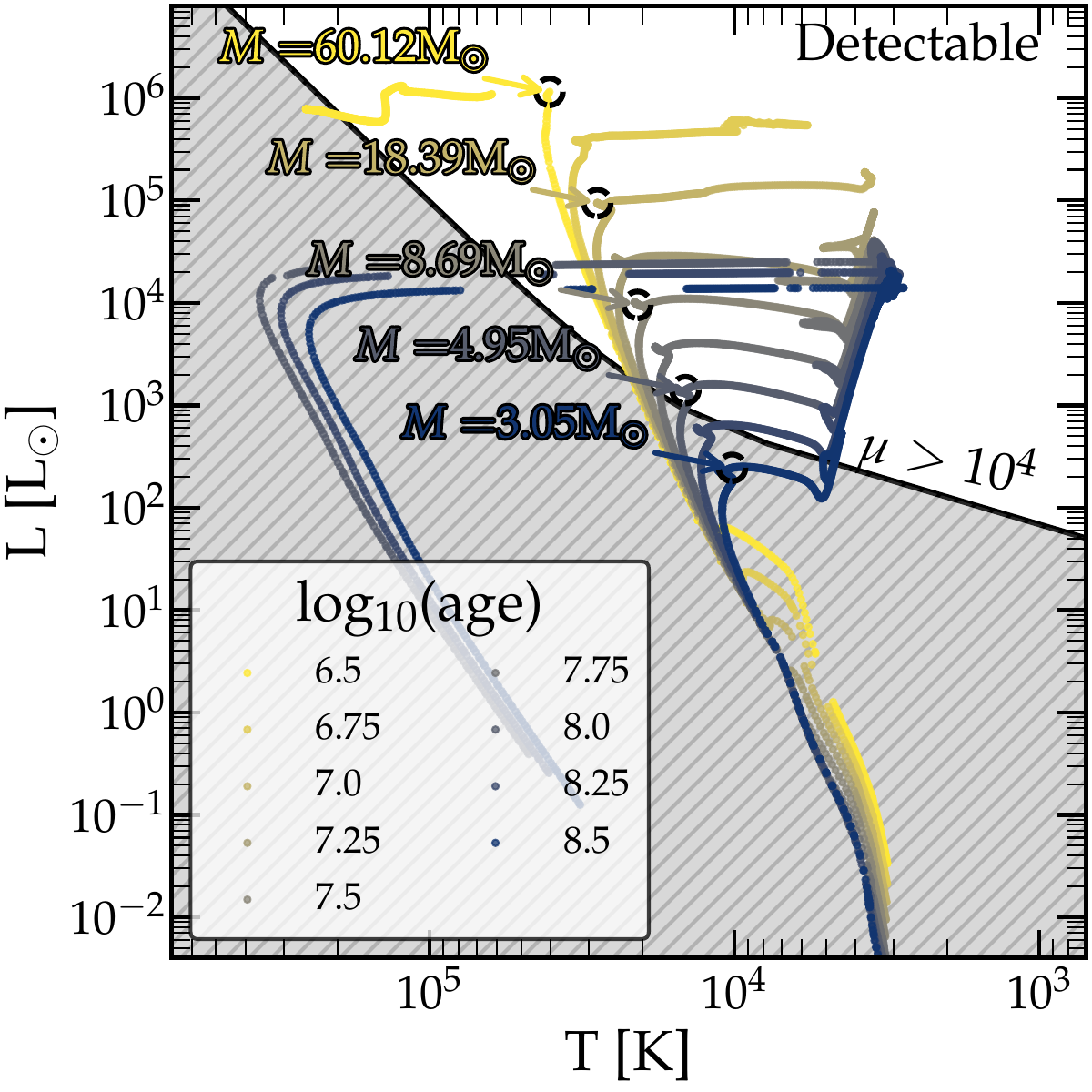}
    \caption{
    Set of isochrones outputted by {\tt FSPS} from the {\it MIST} stellar library at different stellar ages in region 1 and best fit metallicity $Z=0.31Z_\odot$. Each colour represent the same population as it evolves with time with older population growing dimer and redder. The solid black line delimits the regions of the diagram in which a magnification larger than a factor 10$^4$ would be necessary to observe the star in at least one of the eight NIRCam filters used in this work at a limiting magnitude of 30. The shaded region is thus the stars that can not be observed in any filter with these conditions. The arrows mark the turn-off points, where stars leave the main sequence. The corresponding masses indicate the stellar mass at which this transition occurs for a subset the isochrones shown in the plot. 
    }
    \label{fig: Isochrones}
\end{figure}

\subsection{\it Microlensing magnification}
\begin{figure*}[ht!]
    \centering
    \includegraphics[width = \linewidth]{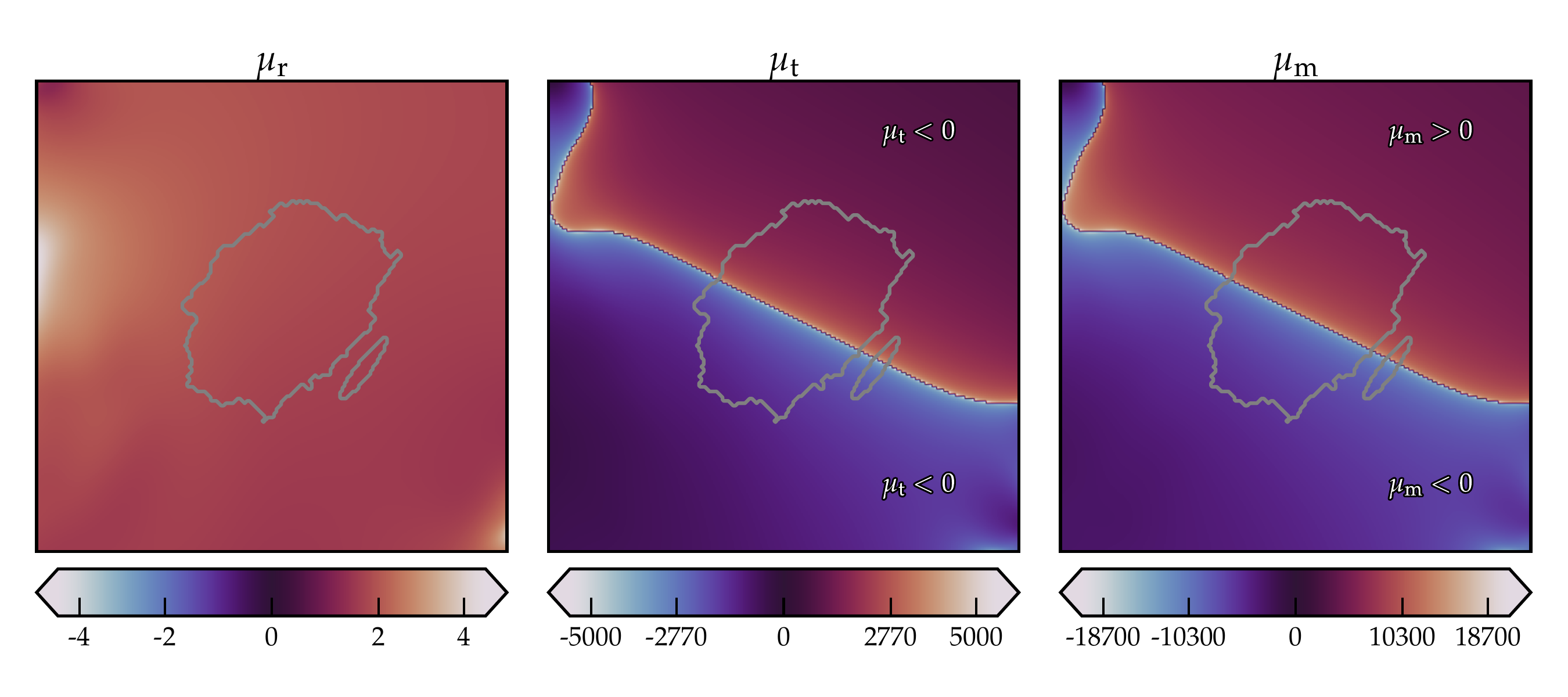}
    \caption{
    Magnification model from WSLAP+ around Warhol. The left panel shows the smooth radial component. The middle panel shows the rapidly varying tangential magnification, with the divergent critical curve near the diagonal. The right panel shows the combined macromodel value, $\mu_m = \mu_t\times\mu_r$. Both $\mu_m$ and $\mu_t$ are shown in log-scale for better visualisation. The region shown is the central 200$\times$200 pixels in Figure 1, enclosed in the yellow square around the Warhol arc. The grey line marks the position of Warhol.
    }
    \label{fig: Lens model}
\end{figure*}

By using {\tt FSPS}, we can not only reconstruct the SFH based on the best-fit parameters from the MCMC, but also generate a set of isochrones using different stellar libraries. In our case, we employ the {\it MIST} library~\citep{Dotter2016, Choi2016, Paxton2011, Paxton2013, Paxton2015}. {\tt FSPS} outputs a file containing various stellar properties, such as age, type, luminosity, temperature, and surface gravity. All stars with the same age form an isochrone. Each star has a weight parameter (fraction by mass), representing the fraction of stars with similar characteristics per solar mass formed. Since we know the total stellar mass formed per pixel and age, we can now determine the total number of stars in each pixel along with their luminosity, temperature, and other required stellar parameters. See figure~\ref{fig: Isochrones} for an output example of {\it FSPS} isochrones.

The final step before incorporating microlensing is to determine the apparent magnitude of each star in the different NIRCam filters. This is achieved using the luminosity and temperature of each star. The apparent magnitude of a source is given by:
\begin{equation}\label{eq: magnitude}
    m = -2.5{\rm log}_{10}\left(F_\nu\right) - 48.6,
\end{equation}
where $F_\nu$ is the flux received in the selected filter, corrected for the filter response $S(\lambda)$, and redshift dependence~\citep{Bessell2012}:
\begin{equation}\label{eq: Flux}
    F_\nu = (1+z)\frac{\int f_{\lambda}\left(\frac{\lambda}{1+z}\right)S(\lambda)\lambda \,d\lambda}{\int S(\lambda)(c/\lambda) \,d\lambda}.
\end{equation}
The specific flux per unit frequency interval is given by 
\begin{equation}\label{eq: flux density}
    f_{\lambda}(\lambda) = \frac{L}{4\pi D_l(z)^2}\frac{{\rm SED}(\lambda)}{\int {\rm SED}(\lambda) \,d\lambda},
\end{equation}
where $L$ is the stellar bolometric luminosity, $D_l(z)$ is the luminosity distance, and the stellar SED per wavelength is represented by a blackbody spectrum without loss of generality. Using equations~(\ref{eq: magnitude}, \ref{eq: Flux}, \ref{eq: flux density}), we can calculate the apparent magnitude of each star in the isochrones. For each star in the isochrone we have their their luminosity as shown in Figure~\ref{fig: Isochrones}. We repeat this for each of the NIRCam filters.

%

Now that we have all the necessary information from the perspective of the sources (i.e., the stars), we can proceed to the second half of the process: microlensing. For this task, we gather information from the macromodel and microlenses, and calculate the magnification statistics. Depending on these elements, a source will experience a flux boost, $\mu_{\rm micro}$, such that its perceived flux is $\mu_{\rm micro}\times F_f(\nu)$, following equation~\ref{eq: magnitude} its new apparent magnitude becomes:
\begin{equation}\label{eq: boost}
    m_{\rm micro} = m -2.5{\rm log}_{10}\left(\mu_{\rm micro}\right).
\end{equation}
$\mu_{\rm micro}$ is drawn from a magnification PDF estimated using {\tt M\_SMiLe}~\citep{Palencia2024}. The PDF varies from pixel to pixel as the macro and microlensing models change. {\tt M\_SMiLe} computes a semi-analytical magnification PDF based on three input elements: tangential macro-magnification $\mu_{\rm t}$, radial macro-magnification $\mu_{\rm r}$, and microlens surface mass density $\Sigma_\ast$. The output at each pixel $i$ is:
\begin{equation}
    p_i({\rm log_{10}}(\mu_{\rm micro}); \mu_{{\rm t}, i}, \mu_{{\rm r}, i}, \Sigma_{\ast, i}),
\end{equation}
which we carefully map to $p_i(\mu)$ and normalise to unity. At first order, this probability density function peaks at $\mu_{\rm m}\,\approx \,\mu_{\rm t}\,\times\,\mu_{\rm r}$, while $\Sigma_{\rm eff}\,=\,\mu_{\rm t}\,\times\,\Sigma_\ast$ controls the width of the PDF. As $\Sigma_{\rm eff}$ increases, higher $\mu_{\rm micro}$ values become more likely, until $\Sigma_{\rm eff}\,\gg\,\Sigma_{\rm crit}$, where the ``more is less" effect begins~\citep{Diego2018,Dai2021,Welch2022a,Palencia2024,Kawai2024}. At this point, extreme values become less likely until the PDF converges to an attractor universal form, and no further differences are observed even at higher $\Sigma_{\rm eff}$, or $\mu_{\rm m}$ 
For a more detailed discussion see~\citep{Diego2018,Palencia2024}.

\begin{figure}[ht!]
    \centering
    \includegraphics[width = \linewidth]{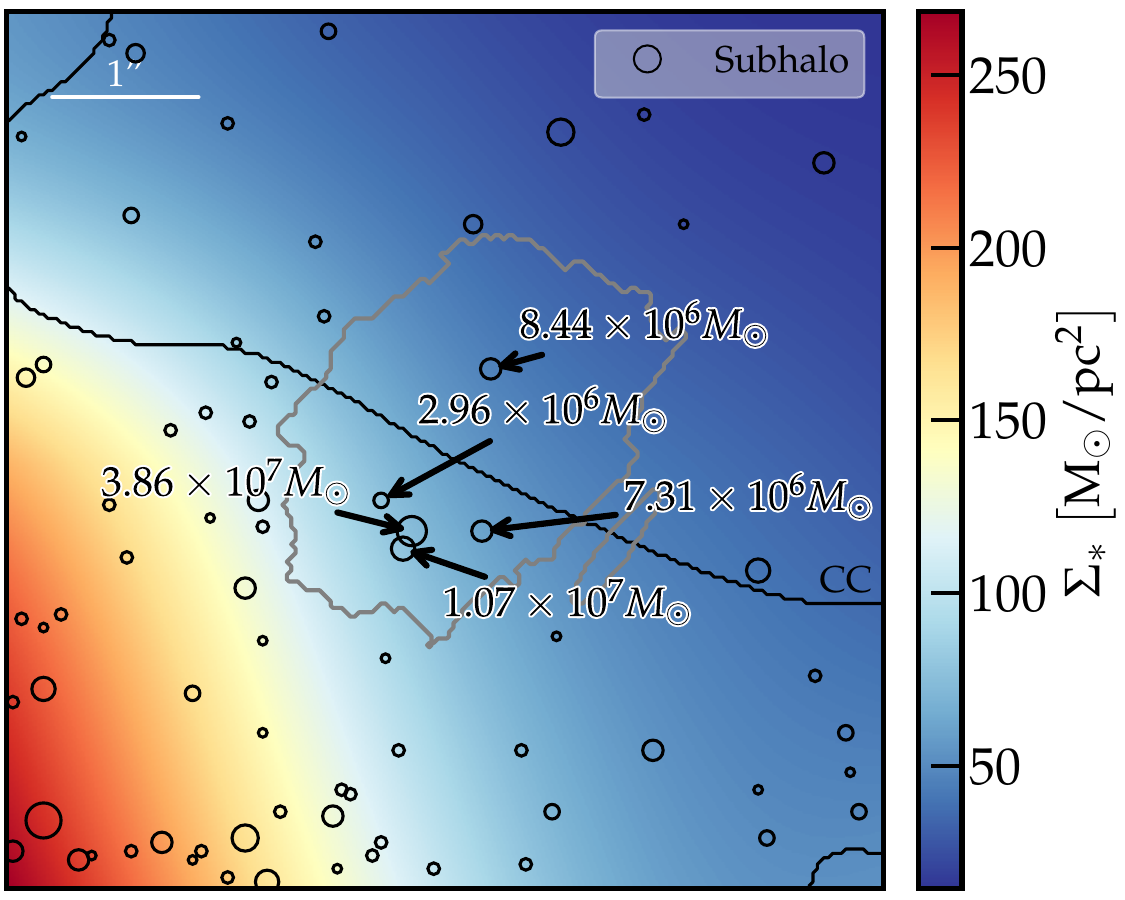}
    \caption{
    $\Sigma_\ast$ modulated by the ICL, relative to a pivotal value of $\sim$60 M$_\odot$/pc$^2$. Black open circles indicate the positions of a random realisation of DM subhalos, with their location probability following the ICL distribution. The subhalos exhibit a mass distribution derived from the small-scale resulting of N-body simulations of dark matter within galaxy clusters at the MACS 0416 redshift. Larger masses are represented by larger circles. The masses of the subhalos intersecting the arc are indicated in the figure. This region shows the central 200×200 pixels in Figure~\ref{fig: MACS 0416}. 
    }
    \label{fig: ICL and GCs}
\end{figure}

We used the WSLAP+ mass distribution $\kappa$ and the shear $\gamma$ for estimating the macrolens model parameters. Both macro-magnification components (see Figure~\ref{fig: Lens model}) can be easily obtained as:
\begin{equation}\label{eq: magnifications}
    \begin{array}{c}
    \mu_{\rm t} = 1/(1-\kappa-\gamma), \\
    \mu_{\rm r} = 1/(1-\kappa+\gamma).
    \end{array}
\end{equation}
$\Sigma_\ast$ is derived from ICL measurements from~\citep{Kaurov2019}. We assume a pivot value of 59.39 M$_\odot$/pc$^2$ modulated according to the ICL variation over the arc as show in Figure~\ref{fig: ICL and GCs}.

\begin{figure*}[ht!]
    \centering
    \includegraphics[width = \linewidth]{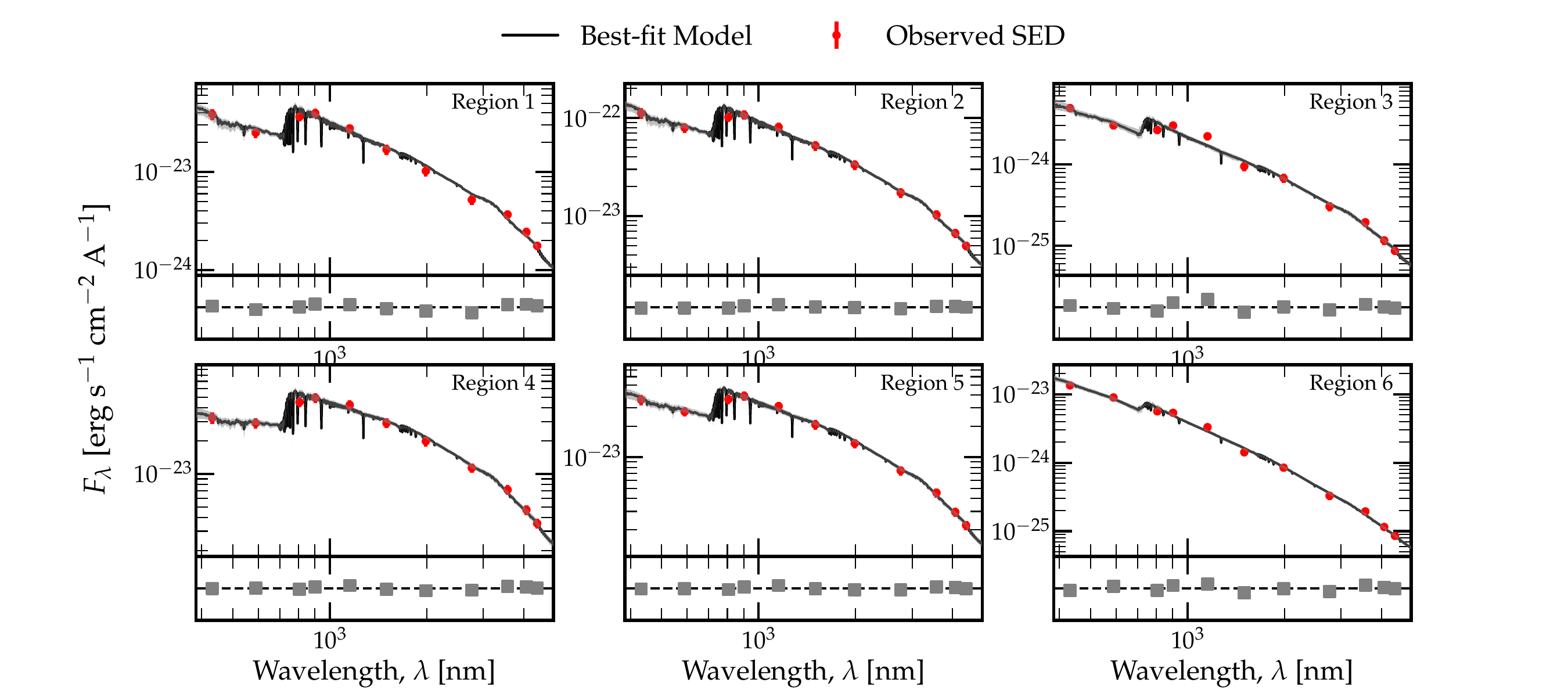}
    \caption{
    SED of Warhol regions and best fit model from the MCMC realizations. The observed data points are showed in red and the solid black line represents the best model from the Flexible Stellar Population Synthesis code. The grey shades area represent the 1$\sigma$ contours. The residuals are showed in the bottom sub-panels for each region. 
    }
    \label{fig: SED fittings}
\end{figure*}

To obtain a more realistic picture of the lensing scheme, we also include GCs as millilenses. As seen in~\citet{Diego2024b}, one can expect $\sim$70 GCs in the 6"$\times$6" area around Warhol. Only a fifth of them will be detectable through photometry, but all of them will locally perturb the macromodel shown in figure~\ref{fig: Lens model}. To achieve this, we randomly place 72 GCs in the 200$\times$200 pixels around Warhol with masses $(10^6\,{\rm M}_\odot\lesssim{\rm M_{milli}}\lesssim10^7\,{\rm M}_\odot)$ 
following the distributions obtained by the MOKA simulations~\citep{Giocoli2012,Giocoli2016} in a galaxy cluster at the MACS 0416 redshift . For each position and mass, we place a Gaussian mass with a full width at half maximum equal to 1 pixel (corresponding to 239 pc at the redshift of the lens), compute the deflection angle of each lens in the pixels of the field of view, add them linearly to those obtained from WSLAP+, and recalculate $\kappa$, $\gamma$, and, following equations~\ref{eq: magnifications}, $\mu_{\rm t}$, and $\mu_{\rm r}$.
In order to investigate the possible impact of compact dark matter (such as primordial black holes), we also assume two additional scenarios, where we include an extra 3$\%$ and 10$\%$ of the total $\kappa$ as microlenses that do not contribute to the ICL, i.e.\ compact dark matter. For reference the critical density in the MACS 0416 and Warhol lensing system is 2970 $M_\odot/{\rm pc}^2$, and $\kappa=\Sigma/\Sigma_{crit}$, where $\Sigma$ is the surface mass density in the lens plane.

\begin{figure*}[ht!]
    \centering
    \includegraphics[width = \linewidth]{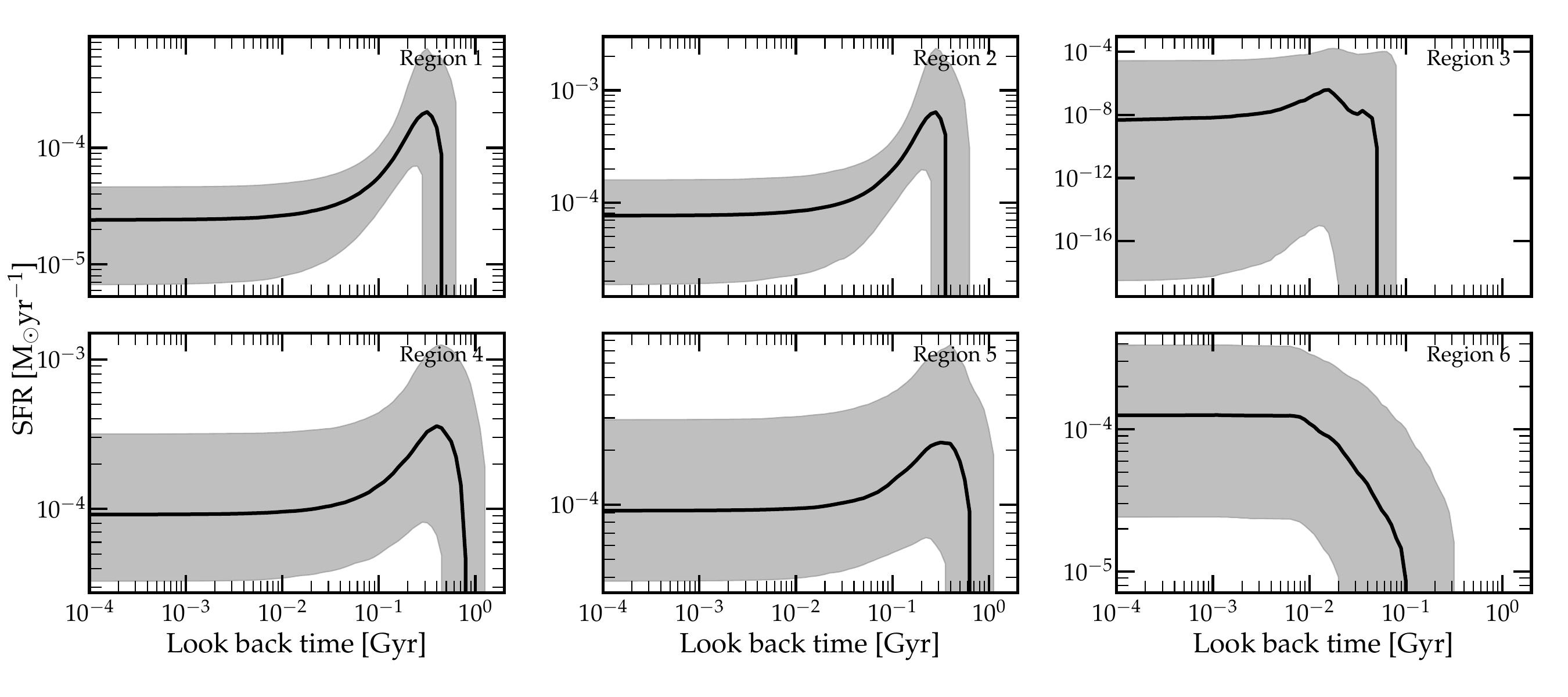}
    \caption{SHFs of Warhol regions from the best fit in our MCMC realizations. The solid black lines represent the median SFR at each lookback time while the grey shaded contour shows the 1$\sigma$ contour.
    }
    \label{fig: sfhs}
\end{figure*}

\begin{figure*}[ht!]
    \centering
    \includegraphics[width = \linewidth]{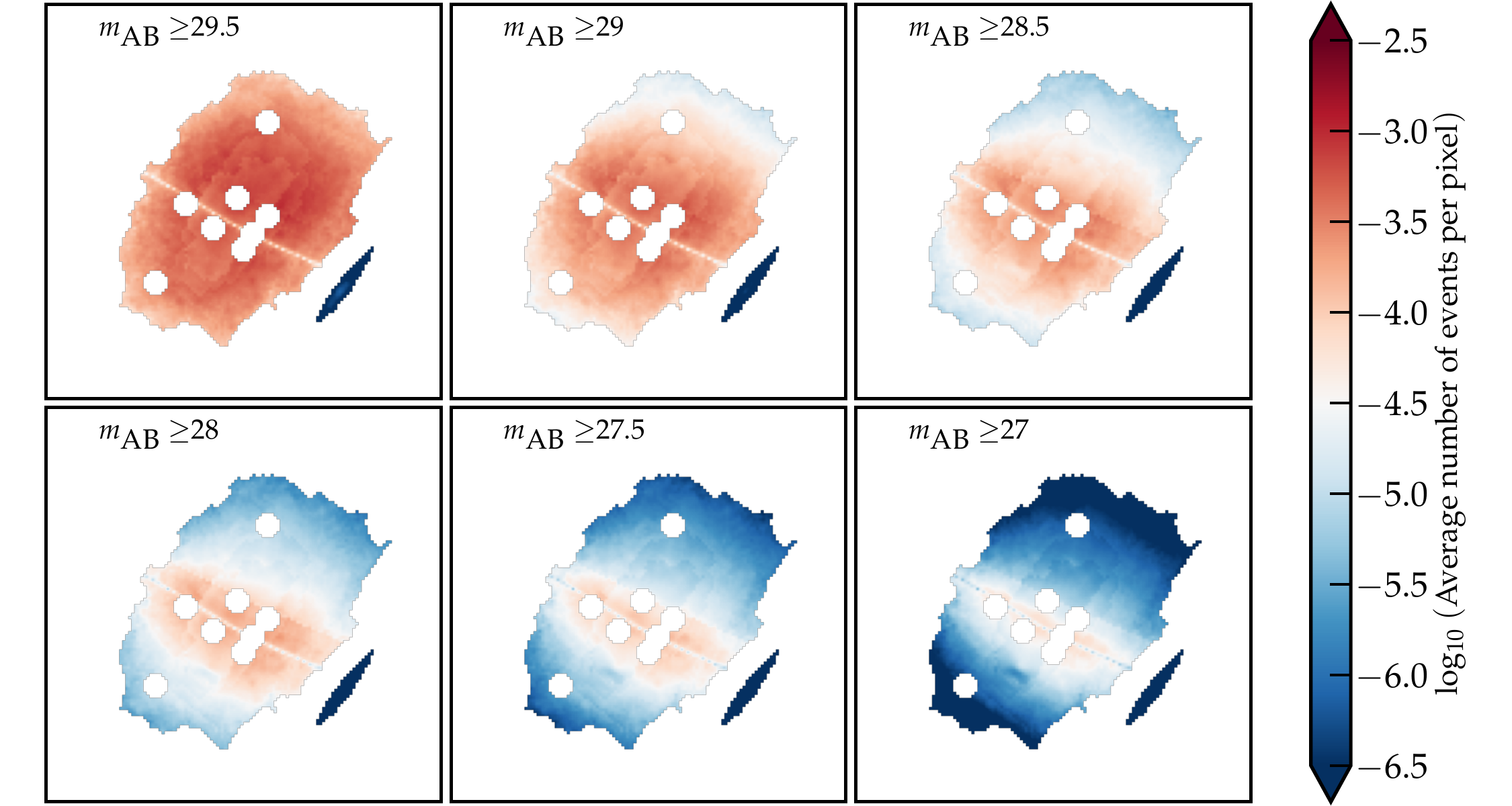}
    \caption{
    Spatial distribution of event probability predicted for different thresholds in F090W.
    Each pixel value represents the expected average number of detections above a certain threshold per pixel.
    The total number of pixels included in the galaxy mask and after masking out the globular clusters and possible events is roughly 5000.
    One pixel length corresponds to 32 mas, or 490 pc on the source plane if $\mu$ = 1.
    In shallow observations we expect to see events only at the critical curve. In deep observations, we expect to see events more uniformly distributed across the arc.
    Lower thresholds predict fewer events, which are asymmetrically distributed, favoring a negative parity. This is due to the increased $\Sigma_{\ast}$ closer to the BCG and the statistical properties of the negative parity regime. As the detection threshold increases, this asymmetry diminishes.}
    \label{fig: N Events spatial}
\end{figure*}

\begin{figure}[ht!]
    \centering
    \includegraphics[width = \linewidth]{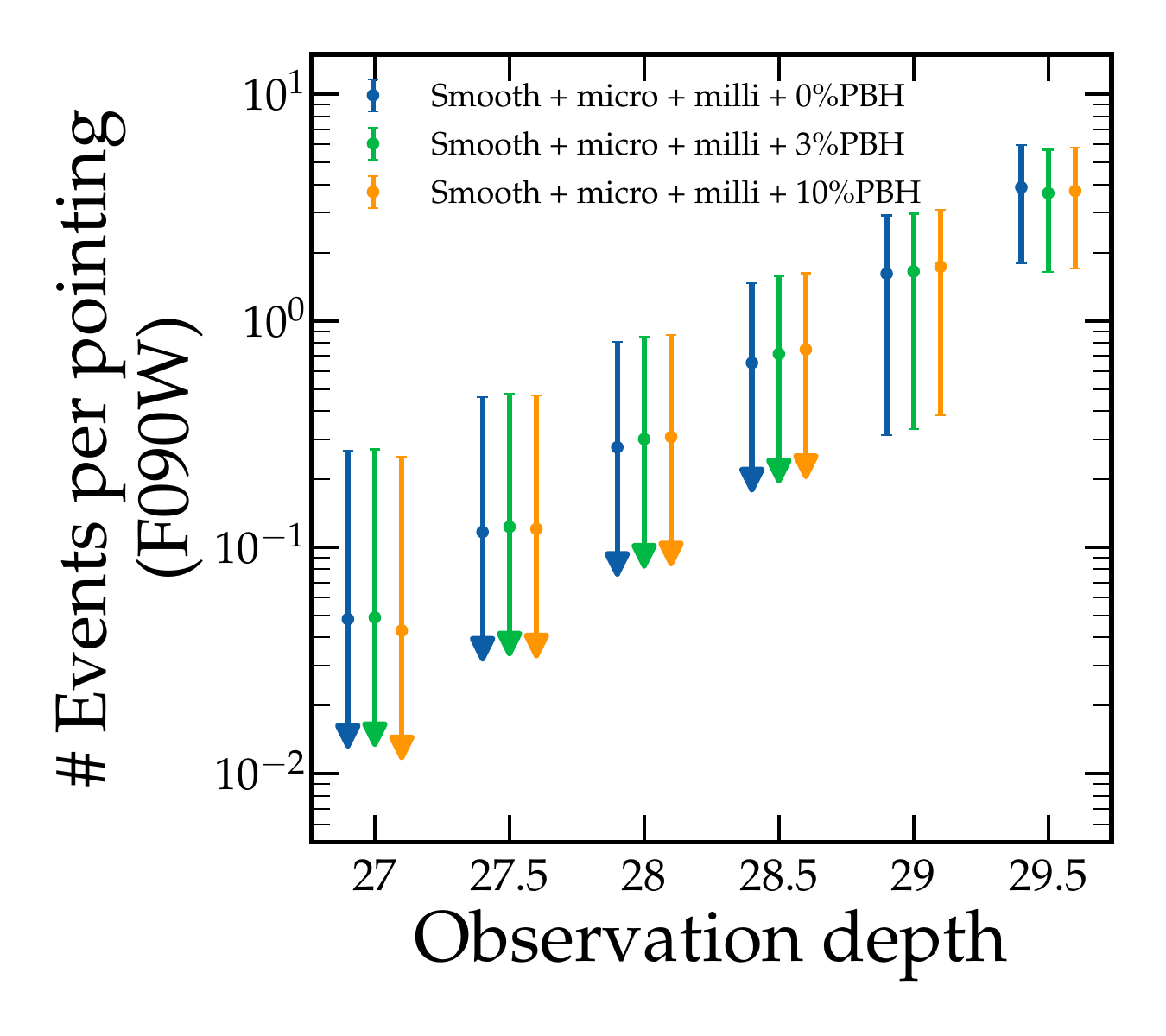}
    \caption{
    Total number of events observed in Warhol for different thresholds in F090W, adding millilenses. Each value is the integrated value of all the pixels in the arc as shown in figure~\ref{fig: N Events spatial}.
    }
    \label{fig: N events integrated}
\end{figure}

By inputting $\mu_{{\rm t}, i}$, $\mu_{{\rm r}, i}$, and $\Sigma_{\ast, i}$ into {\tt M\_SMiLe}, we obtain the magnification probability density function in each pixel that we combined with the estimated number of stars, along with their apparent magnitudes, that we retrieved from our MCMC fitting. Integrating the PDF above a given minimum magnification, set by the star luminosity, dust attenuation, filter of choice, its redshift, and limiting magnitude of the observations, provides the probability of observing a star $j$ in the corresponding pixel $i$:
\begin{equation} \label{eq: p_individual}
P_{i,\,j} = \int_{\mu_{{\rm min},\,j}}^{\infty}p_i(\mu;\,\mu_{{\rm t},i},\, \mu_{{\rm r},i},\, \Sigma_{\ast,i})\,{\rm d}\mu,
\end{equation}
and summing over all stars, weighted by their expected abundance in that pixel, $w_{i,j}$, yields the expected number of events in that pixel:
\begin{equation} \label{eq: p_combined}
n_i = \sum_{j=1}^{N_{\rm star}}P_{i,\,j} w_{i,\,j}.
\end{equation}
The quantity $\mu_{\rm min}$ represents the minimum magnification required for the observed magnitude, following equation~\ref{eq: boost}, to be above a given detection threshold, such that $m_{\rm micro}<m_{\varepsilon}$. Naturally, $\mu_{\rm min}$ depends on the stellar apparent magnitude and the adopted threshold, $\mu_{\rm min}= \mu_{\rm min}(m,\,m_\varepsilon)$.

Summing over all pixels yields the average number of stars detected across the entire arc above the specified threshold:
\begin{equation} \label{eq: integrated N events}
\langle N\rangle = \sum_{i=1}^{N_{\rm pix}} n_i.
\end{equation}
The primary source of uncertainty in this calculation arises from the SFH. The uncertainties in the total stellar mass formed affect the number of stars susceptible to microlensing. By repeating the calculation performed with the median star formation history, but now using the 16th and 84th percentiles obtained from the MCMC chains, we derive a pixel-level uncertainty in the number of events, $\sigma_{n_i}$. The variance in the expected number of events is then given by:
\begin{equation}
\sigma^{2}{\langle N\rangle} = \sqrt{\sum_{i=1}^{N_{\rm pix}}\sigma^{2}_{n_i}}.
\end{equation}
For a given telescope pointing towards the arc with a limiting depth or magnitude threshold, there is an associated Poissonian uncertainty in the expected number of events, which can be approximated as:
\begin{equation}
\sigma_{{\rm Poisson}} = \sqrt{\langle N\rangle}.
\end{equation}
Finally, the total uncertainty in the number of events in the arc is obtained by summing all sources of error in quadrature:
\begin{equation}
\sigma_{{\rm total}} = \sqrt{\sigma^{2}_{\langle N\rangle}+\sigma^{2}_{{\rm Poisson}}}.
\end{equation}

\section{Results} \label{sec:results}
After removing contaminating foregrounds and applying linear cuts in both the PCA dimensions and colour-colour spaces, we identified six distinct regions, as shown in Figure~\ref{fig: Regions}. The second to fourth columns display the normalised SEDs of the pixels, alongside the median SED with the corresponding $1\sigma$ contours.

Figure~\ref{fig: SED fittings} presents the best SED fitting results obtained through MCMC analysis for each region, alongside the residuals for the 11 filters used in the photometric fitting (See appendix~\ref{app: fits} for median values and 1$\sigma$ limits). The extracted SFHs, modelled as exponentially decaying star formation histories, are shown in Figure~\ref{fig: sfhs}. By integrating the SFH over cosmic time, we derive the stellar mass formed in each region at different age bins. This mass is distributed across the region’s pixels by normalising it according to the flux in each pixel, as illustrated in the left panel of Figure~\ref{fig: mass_over_mu}. However, this flux is biased by lensing magnification, necessitating a correction by dividing each pixel’s mass by the macro-magnification predicted at its position within our lens model. The right panel of Figure~\ref{fig: Lens model} displays this magnification, and the corrected stellar mass distribution, free from magnification bias, is shown in the right panel of Figure~\ref{fig: mass_over_mu}.

Summing across all regions, we find that the total stellar mass formed in the multiply lensed region of Warhol over its full SFH is $1.27\pm_{0.03}^{0.41}\times10^6$~M$_\odot$, or half that value if we consider that we have two connected images of the same background galaxy. The oldest stellar population, at 794 Myr (whose SFR is constant for the most recent times), is found in region 4, while the youngest, at 60 Myr, corresponds to region 3. All six regions exhibit subsolar metallicities, with regions 1 and 2 having the highest values at 0.31 and 0.23~Z$_\odot$, respectively, while the remaining regions show metallicities at approximately 10\% of the solar value.

\begin{figure}[ht!]
    \centering
    \includegraphics[width = \linewidth]{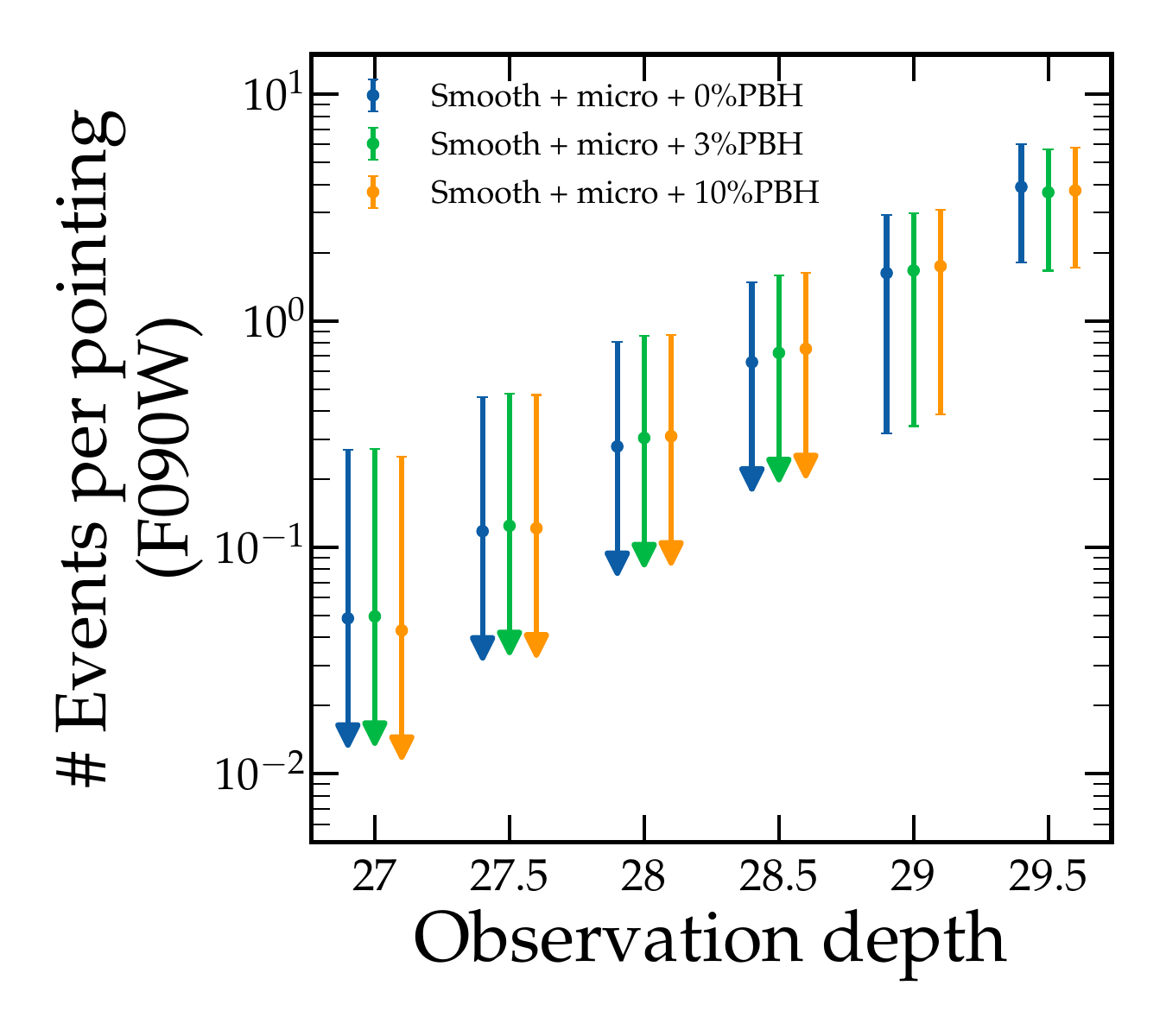}
    \caption{
    Total number of events predicted in Warhol for different thresholds in F090W. Each value is the integrated value of all the pixels in the arc similar to figure~\ref{fig: N Events spatial}.
    }
    \label{fig: N events integrated no millilenses}
\end{figure}

After generating the isochrones corresponding to the stellar populations that best reproduce the observed photometric spectra in Warhol, we applied equation~\ref{eq: p_individual} to each star in the isochrones, weighting them by the total stellar mass formed in that pixel at the corresponding isochrone age. Combining this information via equation~\ref{eq: p_combined}, we obtained the probability density of microlensing events across the Warhol arc at different fiducial observational thresholds, as depicted in Figure~\ref{fig: N Events spatial}. This figure illustrates the distribution of events at various magnitude thresholds in the {\it JWST} NIRCam F090W filter. By integrating the expected event rates across all pixels using equation~\ref{eq: integrated N events}, we derive the expected number of events per pointing in F090W as a function of limiting magnitude, shown in Figure~\ref{fig: N events integrated}.
\begin{figure}[ht!]
    \centering
    \includegraphics[width = \linewidth]{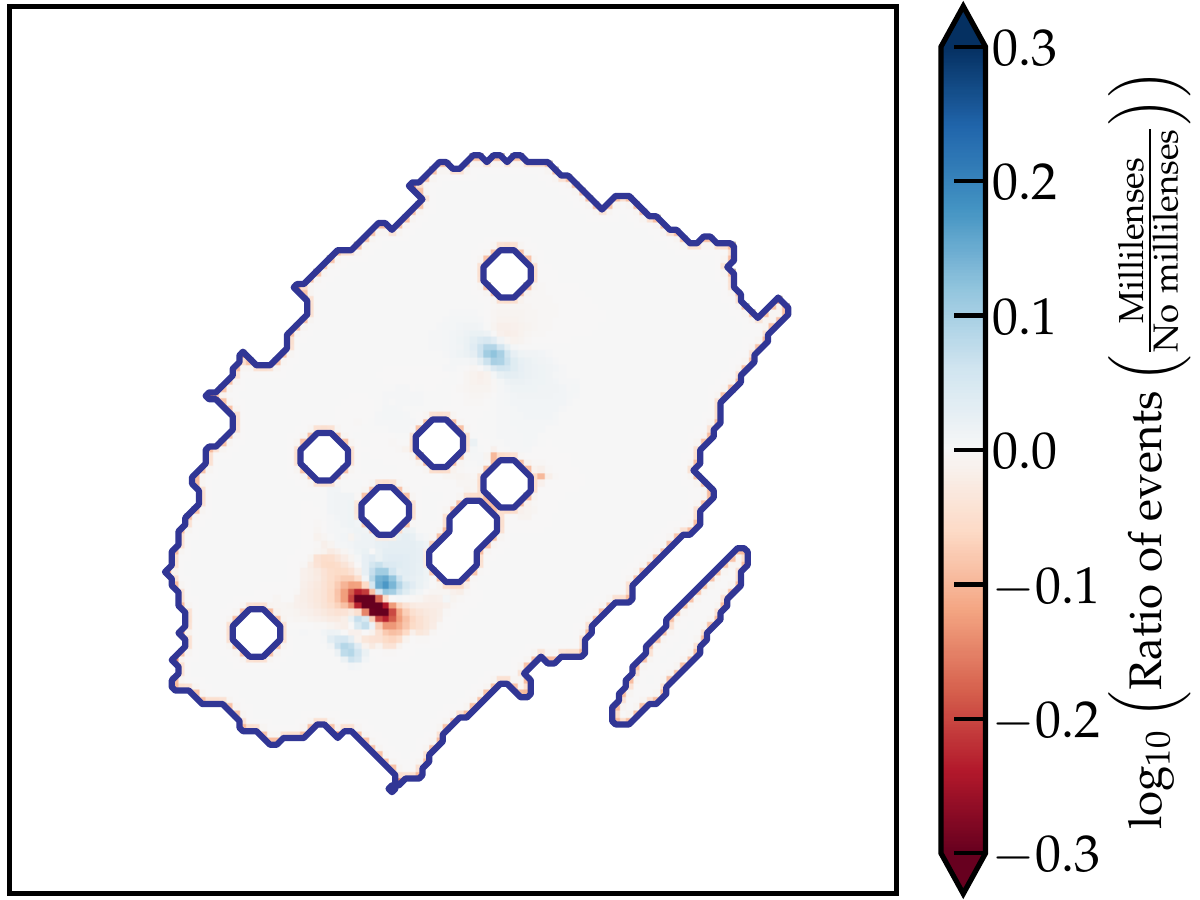}
    \caption{
    Ratio of expected stars magnified above 27.5 magnitudes, assuming millilenses at the positions marked by black circles in Figure~\ref{fig: ICL and GCs}, compared to the model without millilenses. 
    }
    \label{fig: extra millilenses}
\end{figure}
Figures~\ref{fig: N Events spatial} and~\ref{fig: N events integrated} incorporate a lens model that includes millilensing by subhalos, whose mass distribution follows the results obtained from the MOKA simulations~\citep{Giocoli2012,Giocoli2016} in a glaxy cluster at MACS 0416 redshift. The positions of these subhalos are marked by black open circles in Figure~\ref{fig: ICL and GCs}, and their localised impact on the event distribution is evident in Figure~\ref{fig: N Events spatial}, particularly at shallower detection thresholds where their relative influence is greater. Figure~\ref{fig: N events integrated no millilenses} presents the integrated number of expected events per pointing in F090W under the assumption of no millilenses. The overall impact of millilenses on the total event count is at the level of 1\%, well below our uncertainties, which are dominated by Poisson statistics and SFH uncertainties from the SED fitting. However, their effect on the spatial distribution of events is more pronounced, as seen in Figure~\ref{fig: extra millilenses}. This figure shows the ratio of pixel-wise event rates for a 27-magnitude detection threshold in F090W. At the location of a millilens, the event rate is enhanced, whereas the surrounding regions exhibit a localised suppression. This results in a total integrated expectation that remains largely unchanged, as previously seen in Figure~\ref{fig: N events integrated}. In the southern part of the arc, the combined influence of two neighbouring millilenses produces a similar effect to that observed in the northern region, albeit with a more pronounced suppression in the intermediate region between them.

\begin{figure}[ht!]
    \centering
    \includegraphics[width = \linewidth]{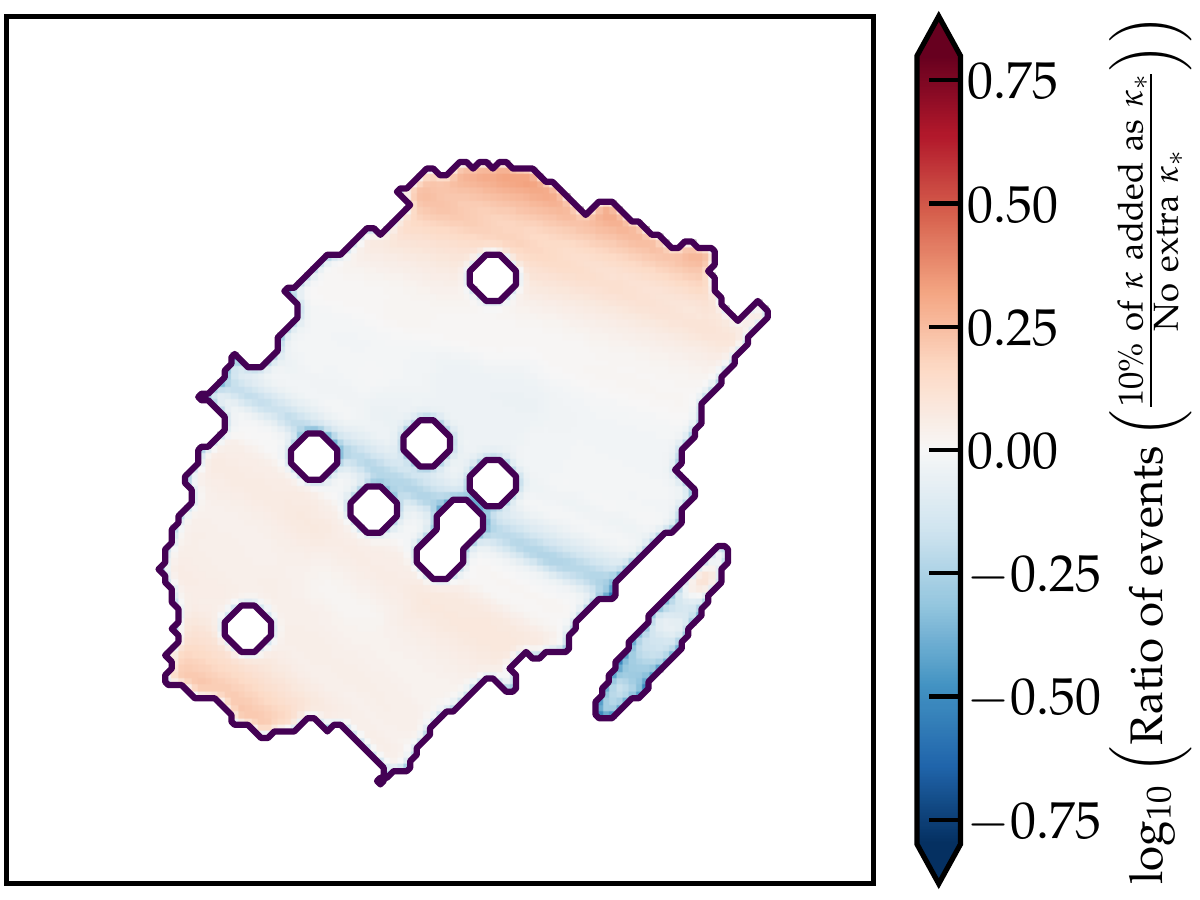}
    \caption{
    Ratio of expected stars magnified above 27.5 magnitudes, increasing $\Sigma_{\ast}$ a 10\% of the total convergence at the position of the arc, compared to the model with the $\Sigma_{\ast}$ derived from the ICL. This mimics the effect of a hypothetical PBH (or other compact dark matter candidate) in the cluster medium. 
    }
    \label{fig: extra kappa}
\end{figure}

The different colours in Figures~\ref{fig: N Events spatial} and~\ref{fig: N events integrated} correspond to varying fractions of microlenses added to the constrained value from the ICL fitting. These additional microlenses, assumed to be compact dark matter objects such as PBHs, are parameterised as a fraction of the smooth surface mass density at the Warhol location. We find that, while the total expected number of events across the arc remains largely unchanged, the slope of the power-law relation between magnitude and event rate is affected. Higher concentrations of compact dark matter result in a steeper slope, indicating an increased sensitivity of event rates to limiting magnitude. 
This can be understood as due to the reduction in maximum magnification near the critical curves as more microlenses (PBHs) are added \citep{Venumadhav2017}. The effect should be compensated in the faint end of the luminosity function, where more events are expected farther away from the critical curve and associated to the increase in number of compact dark matter microlenses. 

This scenario is confirmed when we look at the spatial distribution of events when PBHs are added, which varies with the addition of extra microlenses, as shown in Figure~\ref{fig: extra kappa}, where we plot the event ratio for a scenario with 10\% additional $\kappa$ in microlenses compared to a purely baryonic microlens model. We observe a reduction in the number of events near the critical curve due to the ``more is less" effect~\citep{Diego2018, Palencia2024}, while the event rate is boosted at larger distances. The turnover point in this behaviour depends on the added fraction of microlenses: for an additional 10\% in $\kappa$, we find turnover radii of 1.98$^{\prime\prime}$ and 0.33$^{\prime\prime}$ on the positive and  negative parity sides of the arc, respectively, while for a 3\% increase in $\kappa$, these values shift to 0.96$^{\prime\prime}$ and 0.15$^{\prime\prime}$, approximately half the previous distances.

We conclude that after introducing millilenses into the macromodel and varying the fraction of microlenses in our model, both modifications imprint distinct patterns on the event rate distribution, while their effect on the integrated transient rate remains low or nearly undetectable. Millilenses cluster the event rate probabilities at their location in the lens plane, simultaneously reducing the rate in the surrounding regions. In contrast, increasing the microlens abundance in Warhol, already large due to the baseline value extracted from ICL analysis, further decreases the event rate probability near high magnification regions such as the cluster CC, while enhancing the rates in areas with lower macro magnifications.

We predict an expected number of events per pointing at a limiting magnitude of 29.5 in eight NIRCam filters of order unity (See table~\ref{tab: event rates}). This result is consistent with \citep{Yan2023}, who report seven events in Warhol across four epochs, yielding a mean of 1.75 events per pointing at similar limiting magnitudes. The predicted event rates for all filters and magnitude thresholds are provided in Table~\ref{tab: event rates}. 

\begin{deluxetable*}{c|c|c|c|c|c|c|c|c}
\label{tab: event rates}
\centering
\tabletypesize{\scriptsize}
\tablewidth{0pt} 
\tablecaption{Predicted event rate in the Warhol arc across multiple NIRCam filters and limiting magnitudes.}
\tablehead{ & & & & & & & & \\
Limiting magnitude & F090W & F115W & F150W & F200W & F277W & F356W & F410M & F444W }
\startdata
29.5 & $3.9\pm_{2.1}^{2.1}$ & $5.1\pm_{2.4}^{2.4}$ & $5.6\pm_{2.6}^{2.6}$ & $5.2\pm_{2.4}^{2.5}$ & $4.01\pm_{2.12}^{2.13}$ & $2.7\pm_{1.7}^{1.7}$ & $2.0\pm_{1.5}^{1.5}$ & $1.8\pm_{1.4}^{1.4}$\\
& & & & & & & & \\
29.0 & $1.6\pm_{1.3}^{1.3}$ & $2.1\pm_{1.5}^{1.5}$ & $2.2\pm_{1.5}^{1.5}$ & $2.1\pm_{1.5}^{1.5}$ & $1.6\pm_{1.3}^{1.3}$ & $1.07\pm_{1.05}^{1.05}$ & $0.8\pm_{0.9}^{0.9}$ & $0.8\pm_{0.9}^{0.9}$\\
& & & & & & & & \\
28.5 & $0.7\pm_{0.8}^{0.8}$ & $0.9\pm_{0.9}^{0.9}$ & $0.9\pm_{1.0}^{1.0}$ & $0.8\pm_{0.9}^{0.9}$ & $0.6\pm_{0.8}^{0.8}$ & $0.5\pm_{0.7}^{0.7}$ & $0.4\pm_{0.6}^{0.6}$ & $0.3\pm_{0.6}^{0.6}$\\
& & & & & & & & \\
28.0 & $0.3\pm_{0.5}^{0.5}$ & $0.4\pm_{0.6}^{0.6}$ & $0.4\pm_{0.6}^{0.6}$ & $0.4\pm_{0.6}^{0.6}$ & $0.3\pm_{0.5}^{0.5}$ & $0.2\pm_{0.4}^{0.4}$ & $0.1\pm_{0.4}^{0.4}$ & $0.1\pm_{0.4}^{0.4}$\\
& & & & & & & & \\
27.5 & $0.1\pm_{0.3}^{0.3}$ & $0.2\pm_{0.4}^{0.4}$ & $0.2\pm_{0.4}^{0.4}$ & $0.1\pm_{0.4}^{0.4}$ & $0.1\pm_{0.3}^{0.3}$ & $0.1\pm_{0.3}^{0.3}$ & $0.1\pm_{0.2}^{0.2}$ & $0.1\pm_{0.2}^{0.2}$\\
& & & & & & & & \\
27.0 & $0.04\pm_{0.22}^{0.22}$ & $0.1\pm_{0.3}^{0.3}$ & $0.1\pm_{0.3}^{0.3}$ & $0.1\pm_{0.2}^{0.2}$ & $0.04\pm_{0.21}^{0.21}$ & $0.03\pm_{0.17}^{0.17}$ & $0.02\pm_{0.15}^{0.15}$ & $0.02\pm_{0.14}^{0.14}$\\
\enddata
\end{deluxetable*}

\section{Discussion} \label{sec:discussion}
In this section we discuss the obtained results on the predicted microlensing event rates on eight NIRCam filters on Warhol, a lensed galaxy at redshift $z=0.94$. First we explain the importance of the chosen models, both lensing models and stellar population models. We then evaluate the strengths and weaknesses of our forecasting pipeline.

\subsection{Modelling Effects}\label{ssec: modelling effects}

Our results depend on two key factors: the lens modelling, both macro and micro, and the assumed stellar properties, including priors on stellar population parameters, dust attenuation models, and IMFs. The choice of these models directly impacts the predicted event rates. Here, we discuss our selections and their consequent effects on our results.

\subsubsection{Macrolens}

The choice of macrolens affects our predictions in two primary ways. Firstly, the local values of shear and convergence and their derivatives determine the macromagnification, which serves as an input for our code {\tt M$\_$SMiLe}, influencing the statistical properties of microlensing magnification. Secondly, different models predict slightly varying positions for the critical curve, thereby altering the spatial extent of the parity regimes, which in turn affects the statistical behaviour of micromagnification.

In the case of Warhol, its distinct morphology and bright knots, as shown in figure~\ref{fig: Regions} top left panel, define critical points mirrored in both parity images. The critical curve is precisely bisecting the arc~\citep{Broadhurst2025}, leaving little margin for error in its location at the resolution of our telescopes. Thus, only minor variations in local macromagnification are expected at the arc’s location, implying a minimal impact on the total expected event rate. Moreover, MACS 0416 is one of the best-studied galaxy cluster lenses, hosting the largest number of spectroscopically confirmed sources and hundreds of multiply lensed images. This results in some of the most accurate lens models available for any galaxy cluster. 
However, different lens models may produce the critical curve at the right position yet have different slopes on the potential, affecting the scaling of the magnification with distance to the critical curve, $\mu \propto d^{-1}$. Overall, the entire problem is degenerate in $\Sigma_{\rm eff}=\Sigma_{*}\times\mu$, so uncertainties in the macromodel and the ICL contribution can be reduced to uncertainties in $\Sigma_{\rm eff}$.

However, this level of precision regarding the location of the critical curve in Warhol does not extend to all arcs. For example, Spock, another arc in MACS 0416, is located near a cluster member galaxy and lacks multiple critical points, leading to greater uncertainty in its macrolens model and, consequently, in its predicted event rates~\citep{Li2025}. Other arcs, such as the Dragon Arc in Abell 370, which holds the record for the largest number of detected microlensed stars at cosmological distances~\citep{Fudamoto2024}, are also more challenging to model than Warhol. These examples highlight the necessity for high-quality spectroscopic and photometric data in galaxy clusters to refine mass distribution models and improve constraints on lensing efficiency.

\subsubsection{Micro/Millilens}

The second major component of the lens model is the small-scale perturbers, including both millilenses and microlenses. As illustrated in figure~\ref{fig: extra millilenses}, the spatial distribution of events is influenced by the presence of millilenses. These structures locally enhance the cluster’s lensing efficiency by increasing both the macromagnification and, if the millilens is a globular cluster, the number of microlenses. This results in a higher event rate at the millilens position, albeit at the cost of a slightly reduced rate in the surrounding regions.

Given that tens of globular clusters and possibly DM subhalos are expected to intersect arcs like Warhol, accurately modelling their effects is crucial. Additionally, they can alter the local parity of the macro+milli model, potentially leading to long-duration events in negative parity regions, as is suspected in the case of Mothra~\citep{Diego2023a}. While our study primarily focuses on the detection of events, light curves also play an essential role in a comprehensive analysis of microlensed stars in galaxy clusters.

Besides globular clusters, the actual CDM paradigm predicts the existence of non-luminous millilens-scale halos~\citep{Dai2018, Dai2020b, Williams2024, Ji2025}, which act as millilenses of $10^6-10^8$ M$_\odot$. The spatial distribution of microlensing events can serve as a probe for these small-scale structures as has been demonstrated for the case of ultralight/wave DM~\citep{Broadhurst2025}, particularly when combined with event duration data, although the latter is partially degenerate with the microlens mass. 

Microlenses, a key component in our statistical framework {\tt M$\_$SMiLe}, also play a critical role in shaping the total event rate. A larger microlens population allows for events to occur farther from the critical curve, while simultaneously suppressing events closer to it due to the ``more is less" effect. This results in an overall reduction of event rates at shallower thresholds, whereas deeper pointings exhibit an increase in total expected events. These effects are illustrated in figures~\ref{fig: N events integrated} and~\ref{fig: extra kappa}. We propose leveraging these effects to infer the dark matter abundances at microlens scales in galaxy clusters. This method is expected to be particularly effective in less crowded cluster regions, at higher redshifts where the critical curve moves outward, or in merging clusters such as El Gordo~\citep{Diego2023a,Caminha2023,Frye2023}, where the intracluster medium remains relatively pristine.

\subsubsection{Star Formation History}

The star formation history (SFH) of a galaxy describes how much stellar mass has formed over time. Different galaxies exhibit a variety of SFHs, ranging from continuous star formation to single bursts or multiple episodes of star formation. To model SFHs, we generally adopt one of two approaches: a non-parametric method, where the galaxy’s lifetime is divided into age bins of chosen widths and the total stellar mass formed in each bin is fitted independently, or a parametric approach, where the SFH is assumed to follow a predefined functional form governed by a set of parameters.  

In general, non-parametric SFH modelling is more flexible and offers a better chance of capturing the true star formation history due to a larger number of free parameters. However, having more free parameter (one per age bin) increases the computational complexity, particularly in MCMC-based models, where exploring high-dimensional parameter spaces can hinder convergence, and can also increase the risk of over-fitting.
Parametric models, on the other hand, are less flexible since they impose a predefined SFH shape, but they require fewer parameters, significantly improving the efficiency of MCMC sampling. Nevertheless, incorrect assumptions about the SFH shape can introduce systematic biases in the inferred SFHs.  

In this work, we adopt a simple exponentially decaying SFH, also known as a $\tau$-model, where the star formation rate evolves as:  
\begin{equation}
    \Psi(t) = A\, e^{-\frac{t-t_0}{\tau}}\,\Theta(t-t_0),
\end{equation}
where $t_0$ represents the onset of star formation, $\tau$ controls the decay rate, and $A$ is a normalisation factor.  

Other commonly used parametric SFHs include the delayed exponential decay, double power-law models, constant star formation, multi-burst scenarios, and combinations of these models. The exponentially decaying SFH is a reasonable assumption, as it successfully reproduces the observed SED of many galaxies. 
Similarly, \citet{Li2025} conducted an extensive model comparison for Spock, finding that a non-parametric fit also favoured an exponentially decaying SFH.

We tested different parametric SFH models in each of the six regions of Warhol: constant SFH, delayed exponentially decaying, double powerlaw, and exponentially decaying, finding a preference for the latter model in each region. However, since we did not perform non-parametric modelling or test a combination of parametric models, we may be missing a more complex SFH. It is possible that the true SFH is primarily characterised by an exponential decline but includes short bursts of star formation that are not captured in our analysis. Despite this limitation, analysing discrete regions within Warhol allows us to better account for variations in SFH across the arc.  

Disentangling more complex SFHs would require higher quality data. We conclude that 11 photometric filters provide a reasonable understanding of the SFH in these arcs, but a more detailed approach could be achieved, particularly for the most recent SFH epochs, through spectroscopic analysis.
Emission lines such as H$_\alpha$ and Pa$_\alpha$ are well-known tracers of recent star formation, and future studies should incorporate them to improve the accuracy of SFH reconstruction.

\subsubsection{IMFs}

The stellar initial mass function (IMF) describes the number of main sequence stars formed as a function of stellar mass~\citep{Salpeter1955}. Many fundamental properties of stellar populations, including their SFH, metallicity, luminosity function, and stellar evolution, are directly influenced by the IMF. Measuring the IMF through resolved photometry has only been possible in nearby galaxies such as the Small and Large Magellanic Clouds (SMC and LMC)~\citep{Massey1995, Hunter1995, Hunter1997, Sirianni2000, Sirianni2002,DaRio2009}, M31~\citep{Weisz2015}, and the Milky Way~\citep{Chabrier2003}. In all these cases, the observed IMF is typically modelled as a broken power law, with a commonly accepted slope of 2.3 for high-mass stars ($>1.4M_\odot$) and a shallower slope for lower-mass stars. While the 2.3 slope for massive stars is widely agreed upon, different models vary in their treatment of the low-mass end of the IMF. The most commonly adopted IMFs include those of Kroupa~\citep{Kroupa2001}, Salpeter~\citep{Salpeter1955}, and Chabrier~\citep{Chabrier2003}, among others.  

Transient events typically arise from the brightest members of the stellar populations, including O and B-type main sequence stars, red supergiants (RSGs), and possibly the tip of the red giant branch if the detection threshold is at a sufficiently high magnitude and the distance modulus of the arc is not too large. These stars are not only the most luminous but also the most massive, meaning that transient detections are primarily sensitive to the high-mass tail of the IMF. While lower-mass stars are not directly observed in these events, their presence influences the evolution of nearby massive stars, and due to their shallower IMF slope, they are significantly more abundant.  

Whether the IMF evolves with redshift remains an open question~\citep{Davel2011, Li2023}. Complementary studies, such as \citep{Li2025,Williams+inprep}, have investigated this issue, finding that at $z \sim 1$, a power-law slope of 2.3 is favoured. However, since our focus is on implementing {\tt M$\_$SMiLe} to assess its ability to spatially resolve transient event rate distributions rather than directly constraining the IMF, we adopt a Kroupa IMF~\citep{Kroupa2001} for our analysis. Additionally, introducing the high-mass end slope of the IMF as a free parameter would significantly increase the computational time required for MCMC fitting.

\section{Conclusions} \label{sec:conclusion}

In this work, we have presented a detailed forecast for the detection of highly magnified stars in the Warhol arc at redshift $z=0.94$ lensed by the galaxy cluster MACS 0416, utilising a combination of strong lensing models, stellar population synthesis, and microlensing statistical techniques. Our analysis, based on photometric data from \textit{JWST} and \textit{HST}, provides insights into the spatial distribution and expected number of transient events in different filters and magnitude thresholds.

Our results indicate that the predicted microlensing event rates in the JWST NIRCam filters are in good agreement with current observations~\citep{Yan2023}. 
We have characterised the spatial distribution of microlensing events in Warhol, showing that the highest event rates are not necessarily located at the positions of peak macromagnification, i.e., directly on the CC. Instead, we find that regions with milder macromagnification values report the highest number of expected events. This is consistent with the balance between the increased density of sources at lower magnifications or larger areas in the source plane and the statistical effects of microlensing in highly magnified regions, where the ``more is less" effect suppresses the number of observable transients.

Additionally, we have explored the impact of small-scale structures, including globular clusters, primordial black holes (PBHs), and millilenses, on the spatial distribution of microlensing events. Our results demonstrate that while the total number of detected events remains largely unchanged, the presence of millilenses modifies the clustering of events, enhancing the microlensing efficiency in their immediate vicinity while slightly suppressing it in surrounding regions. This effect is particularly evident at shallow detection thresholds. The presence of dark matter in the form of PBHs and other millilens-scale substructures further alters the spatial pattern of events, creating distinct signatures that could be used to probe the small-scale distribution of dark matter within galaxy clusters.

Despite the limitations inherent to our modelling assumptions, our work demonstrates the power of microlensing as a statistical tool for studying lensed stellar populations at high redshift. Future spectroscopic follow-up observations, particularly targeting emission lines such as H$_\alpha$ and Pa$_\alpha$, will be crucial in refining star formation history models and breaking degeneracies between dust and age effects.

With ongoing and upcoming \textit{JWST} observations, we expect a growing sample of detected lensed stars across multiple galaxy clusters. The methodology developed in this study provides a robust framework for interpreting these discoveries and advancing our understanding of both stellar populations at cosmological distances and the nature of dark matter on sub-galactic scales.

\section*{Achknowledgement}
J.M.P. acknowledges financial support from the Formación de Personal Investigador (FPI) programme, ref.~PRE2020-096261, associated with the Spanish Agencia Estatal de Investigación project MDM-2017-0765-20-2.~J.M.D. acknowledges the support of project PGC2018-101814-B-100 (MCIU/AEI/MINECO/FEDER, UE)~Ministerio de Ciencia, Investigación y Universidades.
B.J.K. acknowledges funding from the Ramón y Cajal Grant RYC2021-034757-I, financed by MCIN/AEI/10.13039/501100011033 and by the European Union “NextGenerationEU"/PRTR. 
S.K.L. acknowledges support from the Research Grants Council (RGC) of Hong Kong through the General Research Fund (GRF) 17312122. 

This research is based on observations made with the NASA/ESA {\it Hubble Space Telescope} and {\it James-Webb Space Telescope} obtained from the Space Telescope Science Institute, which is operated by the Association of Universities for Research in Astronomy, Inc., under NASA contract NAS 5–26555.

We acknowledge Santander Supercomputacion support group at the University of Cantabria who provided access to the supercomputer Altamira Supercomputer at the Institute of Physics
of Cantabria (IFCA-CSIC), member of the Spanish Supercomputing Network, for performing simulations.

\vspace{5mm}
\facilities{HST, JWST}
\software{Python, Numpy, Scipy, FSPS, PixedFit, Matplotlib, SExtractor, Petrofit, Astropy, Photutils, DS9}

\appendix

\section{SED Fitting results}\label{app: fits}

\begin{deluxetable*}{c|c|c|c|c|c}[ht!]
\label{tab: fit_results}
\centering
\tabletypesize{\scriptsize}
\tablewidth{0pt} 
\tablecaption{SED MCMC-fitting results.}
\tablehead{Region & log($M_\ast$) [$M_\odot$] & Z/Z$_\odot$  & Age [Gyr] & $\tau$ & $A_V$}
\startdata
1 &$5.15\pm^{0.07}_{0.07}$ &$0.31\pm^{0.18}_{0.12}$ &$0.44\pm^{0.25}_{0.15}$ &$0.12\pm^{0.11}_{0.06}$ &$0.33\pm^{0.14}_{0.12}$ \\
2 &$5.61\pm^{0.07}_{0.07}$ &$0.23\pm^{0.14}_{0.09}$ &$0.40\pm^{0.25}_{0.14}$ &$0.10\pm^{0.12}_{0.05}$ &$0.38\pm^{0.17}_{0.12}$ \\
3 &$3.8\pm^{0.4}_{0.2}$ &$0.08\pm^{0.12}_{0.06}$ &$0.06\pm^{0.03}_{0.03}$ &$0.001\pm^{0.009}_{0.001}$ &$0.74\pm^{0.8}_{0.5}$ \\
4 &$5.7\pm^{0.2}_{0.1}$ &$0.09\pm^{0.12}_{0.07}$ &$0.8\pm^{0.6}_{0.3}$ &$0.31\pm^{0.30}_{0.16}$ & $0.9\pm^{0.6}_{0.4}$  \\
5 &$5.4\pm^{0.2}_{0.1}$ &$0.08\pm^{0.30}_{0.07}$ &$0.7\pm^{0.5}_{0.3}$ &$0.37\pm^{0.30}_{0.19}$ & $0.8\pm^{0.6}_{0.4}$\\
6 &$4.1\pm^{0.2}_{0.2}$ &$0.07\pm^{0.32}_{0.05}$ &$0.10\pm^{0.23}_{0.08}$ &$0.33\pm^{0.34}_{0.17}$ & $1.0\pm^{0.3}_{0.6}$ \\
\enddata

\tablecomments{The fitted parameters are the median value, and the 1$\sigma$ percentiles.}
\end{deluxetable*}

\bibliography{main}{}

\begin{thebibliography}{}
\expandafter\ifx\csname natexlab\endcsname\relax\def\natexlab#1{#1}\fi
\providecommand{\url}[1]{\href{#1}{#1}}
\providecommand{\dodoi}[1]{doi:~\href{http://doi.org/#1}{\nolinkurl{#1}}}
\providecommand{\doeprint}[1]{\href{http://ascl.net/#1}{\nolinkurl{http://ascl.net/#1}}}
\providecommand{\doarXiv}[1]{\href{https://arxiv.org/abs/#1}{\nolinkurl{https://arxiv.org/abs/#1}}}

\bibitem[{{Abdurro'uf} {et~al.}(2021){Abdurro'uf}, {Lin}, {Wu}, \& {Akiyama}}]{Abdurro'uf2021}
{Abdurro'uf}, {Lin}, Y.-T., {Wu}, P.-F., \& {Akiyama}, M. 2021, \apjs, 254, 15, \dodoi{10.3847/1538-4365/abebe2}

\bibitem[{{Amruth} {et~al.}(2023){Amruth}, {Broadhurst}, {Lim}, {Oguri}, {Smoot}, {Diego}, {Leung}, {Emami}, {Li}, {Chiueh}, {Schive}, {Yeung}, \& {Li}}]{Amruth2023}
{Amruth}, A., {Broadhurst}, T., {Lim}, J., {et~al.} 2023, Nature Astronomy, 7, 736, \dodoi{10.1038/s41550-023-01943-9}

\bibitem[{{Bertin} \& {Arnouts}(1996)}]{Bertin1996}
{Bertin}, E., \& {Arnouts}, S. 1996, \aaps, 117, 393, \dodoi{10.1051/aas:1996164}

\bibitem[{{Bessell} \& {Murphy}(2012)}]{Bessell2012}
{Bessell}, M., \& {Murphy}, S. 2012, \pasp, 124, 140, \dodoi{10.1086/664083}

\bibitem[{Bradley {et~al.}(2024)Bradley, Sipőcz, Robitaille, Tollerud, Vinícius, Deil, Barbary, Wilson, Busko, Donath, Günther, Cara, Lim, Meßlinger, Burnett, Conseil, Droettboom, Bostroem, Bray, Bratholm, Jamieson, Ginsburg, Barentsen, Craig, Pascual, Rathi, Perrin, Morris, \& Perren}]{Bradley2024}
Bradley, L., Sipőcz, B., Robitaille, T., {et~al.} 2024, Zenodo, \dodoi{10.5281/zenodo.12585239}

\bibitem[{{Broadhurst} {et~al.}(2025){Broadhurst}, {Li}, {Alfred}, {Diego}, {Morilla}, {Kelly}, {Sun}, {Oguri}, {Williams}, {Windhorst}, {Zitrin}, {Abe}, {Chen}, {Dai}, {Fudamoto}, {Kawai}, {Lim}, {Liu}, {Meena}, {Palencia}, {Smoot}, \& {Williams}}]{Broadhurst2025}
{Broadhurst}, T., {Li}, S.~K., {Alfred}, A., {et~al.} 2025, \apjl, 978, L5, \dodoi{10.3847/2041-8213/ad9aa8}

\bibitem[{{Caminha} {et~al.}(2023){Caminha}, {Grillo}, {Rosati}, {Liu}, {Acebron}, {Bergamini}, {Caputi}, {Mercurio}, {Tozzi}, {Vanzella}, {Demarco}, {Frye}, {Rosani}, \& {Sharon}}]{Caminha2023}
{Caminha}, G.~B., {Grillo}, C., {Rosati}, P., {et~al.} 2023, \aap, 678, A3, \dodoi{10.1051/0004-6361/202244897}

\bibitem[{{Carr} \& {K{\"u}hnel}(2020)}]{Carr2020}
{Carr}, B., \& {K{\"u}hnel}, F. 2020, Annual Review of Nuclear and Particle Science, 70, 355, \dodoi{10.1146/annurev-nucl-050520-125911}

\bibitem[{{Chabrier}(2003)}]{Chabrier2003}
{Chabrier}, G. 2003, \pasp, 115, 763, \dodoi{10.1086/376392}

\bibitem[{{Chen} {et~al.}(2019){Chen}, {Kelly}, {Diego}, {Oguri}, {Williams}, {Zitrin}, {Treu}, {Smith}, {Broadhurst}, {Kaiser}, {Foley}, {Filippenko}, {Salo}, {Hjorth}, \& {Selsing}}]{Chen2019}
{Chen}, W., {Kelly}, P.~L., {Diego}, J.~M., {et~al.} 2019, \apj, 881, 8, \dodoi{10.3847/1538-4357/ab297d}

\bibitem[{{Choi} {et~al.}(2016){Choi}, {Dotter}, {Conroy}, {Cantiello}, {Paxton}, \& {Johnson}}]{Choi2016}
{Choi}, J., {Dotter}, A., {Conroy}, C., {et~al.} 2016, \apj, 823, 102, \dodoi{10.3847/0004-637X/823/2/102}

\bibitem[{{Conroy} \& {Gunn}(2010)}]{Conroy2010b}
{Conroy}, C., \& {Gunn}, J.~E. 2010, \apj, 712, 833, \dodoi{10.1088/0004-637X/712/2/833}

\bibitem[{{Conroy} {et~al.}(2009){Conroy}, {Gunn}, \& {White}}]{Conroy2009}
{Conroy}, C., {Gunn}, J.~E., \& {White}, M. 2009, \apj, 699, 486, \dodoi{10.1088/0004-637X/699/1/486}

\bibitem[{{Conroy} {et~al.}(2010){Conroy}, {White}, \& {Gunn}}]{Conroy2010a}
{Conroy}, C., {White}, M., \& {Gunn}, J.~E. 2010, \apj, 708, 58, \dodoi{10.1088/0004-637X/708/1/58}

\bibitem[{{Da Rio} {et~al.}(2009){Da Rio}, {Gouliermis}, \& {Henning}}]{DaRio2009}
{Da Rio}, N., {Gouliermis}, D.~A., \& {Henning}, T. 2009, \apj, 696, 528, \dodoi{10.1088/0004-637X/696/1/528}

\bibitem[{{Dai} \& {Miralda-Escud{\'e}}(2020)}]{Dai2020a}
{Dai}, L., \& {Miralda-Escud{\'e}}, J. 2020, \aj, 159, 49, \dodoi{10.3847/1538-3881/ab5e83}

\bibitem[{{Dai} \& {Pascale}(2021)}]{Dai2021}
{Dai}, L., \& {Pascale}, M. 2021, arXiv e-prints, arXiv:2104.12009, \dodoi{10.48550/arXiv.2104.12009}

\bibitem[{{Dai} {et~al.}(2018){Dai}, {Venumadhav}, {Kaurov}, \& {Miralda-Escud}}]{Dai2018}
{Dai}, L., {Venumadhav}, T., {Kaurov}, A.~A., \& {Miralda-Escud}, J. 2018, \apj, 867, 24, \dodoi{10.3847/1538-4357/aae478}

\bibitem[{{Dai} {et~al.}(2020){Dai}, {Kaurov}, {Sharon}, {Florian}, {Miralda-Escud{\'e}}, {Venumadhav}, {Frye}, {Rigby}, \& {Bayliss}}]{Dai2020b}
{Dai}, L., {Kaurov}, A.~A., {Sharon}, K., {et~al.} 2020, \mnras, 495, 3192, \dodoi{10.1093/mnras/staa1355}

\bibitem[{{Dav{\'e}}(2011)}]{Davel2011}
{Dav{\'e}}, R. 2011, in Astronomical Society of the Pacific Conference Series, Vol. 440, UP2010: Have Observations Revealed a Variable Upper End of the Initial Mass Function?, ed. M.~{Treyer}, T.~{Wyder}, J.~{Neill}, M.~{Seibert}, \& J.~{Lee}, 353, \dodoi{10.48550/arXiv.1008.5283}

\bibitem[{{Diego} {et~al.}(2005){Diego}, {Protopapas}, {Sandvik}, \& {Tegmark}}]{Diego2005}
{Diego}, J.~M., {Protopapas}, P., {Sandvik}, H.~B., \& {Tegmark}, M. 2005, \mnras, 360, 477, \dodoi{10.1111/j.1365-2966.2005.09021.x}

\bibitem[{{Diego} {et~al.}(2007){Diego}, {Tegmark}, {Protopapas}, \& {Sandvik}}]{Diego2007}
{Diego}, J.~M., {Tegmark}, M., {Protopapas}, P., \& {Sandvik}, H.~B. 2007, \mnras, 375, 958, \dodoi{10.1111/j.1365-2966.2007.11380.x}

\bibitem[{{Diego} {et~al.}(2018){Diego}, {Kaiser}, {Broadhurst}, {Kelly}, {Rodney}, {Morishita}, {Oguri}, {Ross}, {Zitrin}, {Jauzac}, {Richard}, {Williams}, {Vega-Ferrero}, {Frye}, \& {Filippenko}}]{Diego2018}
{Diego}, J.~M., {Kaiser}, N., {Broadhurst}, T., {et~al.} 2018, \apj, 857, 25.
\newblock \doarXiv{1706.10281}

\bibitem[{{Diego} {et~al.}(2023{\natexlab{a}}){Diego}, {Sun}, {Yan}, {Furtak}, {Zackrisson}, {Dai}, {Kelly}, {Nonino}, {Adams}, {Meena}, {Willner}, {Zitrin}, {Cohen}, {D'Silva}, {Jansen}, {Summers}, {Windhorst}, {Coe}, {Conselice}, {Driver}, {Frye}, {Grogin}, {Koekemoer}, {Marshall}, {Pirzkal}, {Robotham}, {Rutkowski}, {Ryan}, {Tompkins}, {Willmer}, \& {Bhatawdekar}}]{Diego2023b}
{Diego}, J.~M., {Sun}, B., {Yan}, H., {et~al.} 2023{\natexlab{a}}, \aap, 679, A31, \dodoi{10.1051/0004-6361/202347556}

\bibitem[{{Diego} {et~al.}(2023{\natexlab{b}}){Diego}, {Meena}, {Adams}, {Broadhurst}, {Dai}, {Coe}, {Frye}, {Kelly}, {Koekemoer}, {Pascale}, {Willner}, {Zackrisson}, {Zitrin}, {Windhorst}, {Cohen}, {Jansen}, {Summers}, {Tompkins}, {Conselice}, {Driver}, {Yan}, {Grogin}, {Marshall}, {Pirzkal}, {Robotham}, {Ryan}, {Willmer}, {Bradley}, {Caminha}, {Caputi}, {Carleton}, \& {Kamieneski}}]{Diego2023a}
{Diego}, J.~M., {Meena}, A.~K., {Adams}, N.~J., {et~al.} 2023{\natexlab{b}}, \aap, 672, A3, \dodoi{10.1051/0004-6361/202245238}

\bibitem[{{Diego} {et~al.}(2024{\natexlab{a}}){Diego}, {Li}, {Meena}, {Niemiec}, {Acebron}, {Jauzac}, {Struble}, {Amruth}, {Broadhurst}, {Cerny}, {Ebeling}, {Filippenko}, {Jullo}, {Kelly}, {Koekemoer}, {Lagattuta}, {Lim}, {Limousin}, {Mahler}, {Patel}, {Remolina}, {Richard}, {Sharon}, {Steinhardt}, {Umetsu}, {Williams}, {Zitrin}, {Palencia}, {Dai}, {Ji}, \& {Pascale}}]{Diego2024a}
{Diego}, J.~M., {Li}, S.~K., {Meena}, A.~K., {et~al.} 2024{\natexlab{a}}, \aap, 681, A124, \dodoi{10.1051/0004-6361/202346761}

\bibitem[{{Diego} {et~al.}(2024{\natexlab{b}}){Diego}, {Adams}, {Willner}, {Harvey}, {Broadhurst}, {Cohen}, {Jansen}, {Summers}, {Windhorst}, {D'Silva}, {Koekemoer}, {Coe}, {Conselice}, {Driver}, {Frye}, {Grogin}, {Marshall}, {Nonino}, {Ortiz}, {Pirzkal}, {Robotham}, {Ryan}, {Willmer}, {Yan}, {Sun}, {Hainline}, {Berkheimer}, {Polletta}, \& {Zitrin}}]{Diego2024b}
{Diego}, J.~M., {Adams}, N.~J., {Willner}, S.~P., {et~al.} 2024{\natexlab{b}}, \aap, 690, A114, \dodoi{10.1051/0004-6361/202349119}

\bibitem[{{Dotter}(2016)}]{Dotter2016}
{Dotter}, A. 2016, \apjs, 222, 8, \dodoi{10.3847/0067-0049/222/1/8}

\bibitem[{{Ebeling} {et~al.}(2001){Ebeling}, {Edge}, \& {Henry}}]{Ebeling2001}
{Ebeling}, H., {Edge}, A.~C., \& {Henry}, J.~P. 2001, \apj, 553, 668, \dodoi{10.1086/320958}

\bibitem[{{Foreman-Mackey} {et~al.}(2013){Foreman-Mackey}, {Hogg}, {Lang}, \& {Goodman}}]{Foreman-Mackey2013}
{Foreman-Mackey}, D., {Hogg}, D.~W., {Lang}, D., \& {Goodman}, J. 2013, \pasp, 125, 306, \dodoi{10.1086/670067}

\bibitem[{{Frye} {et~al.}(2023){Frye}, {Pascale}, {Foo}, {Leimbach}, {Garuda}, {Robles}, {Summers}, {Diaz}, {Kamieneski}, {Furtak}, {Cohen}, {Diego}, {Beauchesne}, {Windhorst}, {Willner}, {Koekemoer}, {Zitrin}, {Caminha}, {Caputi}, {Coe}, {Conselice}, {Dai}, {Dole}, {Driver}, {Grogin}, {Harrington}, {Jansen}, {Kneib}, {Lehnert}, {Lowenthal}, {Marshall}, {Menanteau}, {Pampliega}, {Pirzkal}, {Polletta}, {Richard}, {Robotham}, {Ryan}, {Rutkowski}, {Sif{\'o}n}, {Tompkins}, {Wang}, {Yan}, \& {Yun}}]{Frye2023}
{Frye}, B.~L., {Pascale}, M., {Foo}, N., {et~al.} 2023, \apj, 952, 81, \dodoi{10.3847/1538-4357/acd929}

\bibitem[{{Fudamoto} {et~al.}(2024){Fudamoto}, {Sun}, {Diego}, {Dai}, {Oguri}, {Zitrin}, {Zackrisson}, {Jauzac}, {Lagattuta}, {Egami}, {Iani}, {Windhorst}, {Abe}, {Bauer}, {Bian}, {Bhatawdekar}, {Broadhurst}, {Cai}, {Chen}, {Chen}, {Cohen}, {Conselice}, {Espada}, {Foo}, {Frye}, {Fujimoto}, {Furtak}, {Golubchik}, {Hsiao}, {Jolly}, {Kawai}, {Kelly}, {Koekemoer}, {Kohno}, {Kokorev}, {Li}, {Li}, {Lin}, {Magdis}, {Meena}, {Nabizadeh}, {Richard}, {Steinhardt}, {Wu}, {Zhu}, \& {Zou}}]{Fudamoto2024}
{Fudamoto}, Y., {Sun}, F., {Diego}, J.~M., {et~al.} 2024, arXiv e-prints, arXiv:2404.08045, \dodoi{10.48550/arXiv.2404.08045}

\bibitem[{{Geda} {et~al.}(2022){Geda}, {Crawford}, {Hunt}, {Bershady}, {Tollerud}, \& {Randriamampandry}}]{Geda2022}
{Geda}, R., {Crawford}, S.~M., {Hunt}, L., {et~al.} 2022, \aj, 163, 202, \dodoi{10.3847/1538-3881/ac5908}

\bibitem[{{Giocoli} {et~al.}(2016){Giocoli}, {Bonamigo}, {Limousin}, {Meneghetti}, {Moscardini}, {Angulo}, {Despali}, \& {Jullo}}]{Giocoli2016}
{Giocoli}, C., {Bonamigo}, M., {Limousin}, M., {et~al.} 2016, \mnras, 462, 167, \dodoi{10.1093/mnras/stw1651}

\bibitem[{{Giocoli} {et~al.}(2012){Giocoli}, {Meneghetti}, {Bartelmann}, {Moscardini}, \& {Boldrin}}]{Giocoli2012}
{Giocoli}, C., {Meneghetti}, M., {Bartelmann}, M., {Moscardini}, L., \& {Boldrin}, M. 2012, \mnras, 421, 3343, \dodoi{10.1111/j.1365-2966.2012.20558.x}

\bibitem[{{Goodman} \& {Weare}(2010)}]{Goodman2010}
{Goodman}, J., \& {Weare}, J. 2010, Communications in Applied Mathematics and Computational Science, 5, 65, \dodoi{10.2140/camcos.2010.5.65}

\bibitem[{{Green} \& {Kavanagh}(2021)}]{Green2021}
{Green}, A.~M., \& {Kavanagh}, B.~J. 2021, Journal of Physics G Nuclear Physics, 48, 043001, \dodoi{10.1088/1361-6471/abc534}

\bibitem[{{Hunter} {et~al.}(1997){Hunter}, {Light}, {Holtzman}, \& {Grillmair}}]{Hunter1997}
{Hunter}, D.~A., {Light}, R.~M., {Holtzman}, J.~A., \& {Grillmair}, C.~J. 1997, \apj, 478, 124, \dodoi{10.1086/303790}

\bibitem[{{Hunter} {et~al.}(1995){Hunter}, {Shaya}, {Holtzman}, {Light}, {O'Neil}, \& {Lynds}}]{Hunter1995}
{Hunter}, D.~A., {Shaya}, E.~J., {Holtzman}, J.~A., {et~al.} 1995, \apj, 448, 179, \dodoi{10.1086/175950}

\bibitem[{{Ji} \& {Dai}(2025)}]{Ji2025}
{Ji}, L., \& {Dai}, L. 2025, \apj, 980, 190, \dodoi{10.3847/1538-4357/ada76a}

\bibitem[{Johnson {et~al.}(2024)Johnson, Foreman-Mackey, Sick, Leja, Walmsley, Tollerud, Leung, Scott, \& Park}]{pythonFSPS}
Johnson, B., Foreman-Mackey, D., Sick, J., {et~al.} 2024, Zenodo, \dodoi{10.5281/zenodo.12447779}

\bibitem[{{Kaurov} {et~al.}(2019){Kaurov}, {Dai}, {Venumadhav}, {Miralda-Escud{\'e}}, \& {Frye}}]{Kaurov2019}
{Kaurov}, A.~A., {Dai}, L., {Venumadhav}, T., {Miralda-Escud{\'e}}, J., \& {Frye}, B. 2019, \apj, 880, 58, \dodoi{10.3847/1538-4357/ab2888}

\bibitem[{{Kawai} \& {Oguri}(2024)}]{Kawai2024}
{Kawai}, H., \& {Oguri}, M. 2024, \prd, 110, 083514, \dodoi{10.1103/PhysRevD.110.083514}

\bibitem[{{Kelly} {et~al.}(2018){Kelly}, {Diego}, {Rodney}, {Kaiser}, {Broadhurst}, {Zitrin}, {Treu}, {P{\'e}rez-Gonz{\'a}lez}, {Morishita}, {Jauzac}, {Selsing}, {Oguri}, {Pueyo}, {Ross}, {Filippenko}, {Smith}, {Hjorth}, {Cenko}, {Wang}, {Howell}, {Richard}, {Frye}, {Jha}, {Foley}, {Norman}, {Bradac}, {Zheng}, {Brammer}, {Benito}, {Cava}, {Christensen}, {de Mink}, {Graur}, {Grillo}, {Kawamata}, {Kneib}, {Matheson}, {McCully}, {Nonino}, {P{\'e}rez-Fournon}, {Riess}, {Rosati}, {Schmidt}, {Sharon}, \& {Weiner}}]{Kelly2018}
{Kelly}, P.~L., {Diego}, J.~M., {Rodney}, S., {et~al.} 2018, Nature Astronomy, 2, 334.
\newblock \doarXiv{1706.10279}

\bibitem[{{Kelly} {et~al.}(2022){Kelly}, {Chen}, {Alfred}, {Broadhurst}, {Diego}, {Emami}, {Filippenko}, {Keen}, {Li}, {Lim}, {Meena}, {Oguri}, {Scarlata}, {Treu}, {Williams}, {Williams}, {Zhou}, {Zitrin}, {Foley}, {Jha}, {Kaiser}, {Mehta}, {Rieck}, {Salo}, {Smith}, \& {Weisz}}]{Kelly2022}
{Kelly}, P.~L., {Chen}, W., {Alfred}, A., {et~al.} 2022, arXiv e-prints, arXiv:2211.02670, \dodoi{10.48550/arXiv.2211.02670}

\bibitem[{{Koekemoer} {et~al.}(2011){Koekemoer}, {Faber}, {Ferguson}, {Grogin}, {Kocevski}, {Koo}, {Lai}, {Lotz}, {Lucas}, {McGrath}, {Ogaz}, {Rajan}, {Riess}, {Rodney}, {Strolger}, {Casertano}, {Castellano}, {Dahlen}, {Dickinson}, {Dolch}, {Fontana}, {Giavalisco}, {Grazian}, {Guo}, {Hathi}, {Huang}, {van der Wel}, {Yan}, {Acquaviva}, {Alexander}, {Almaini}, {Ashby}, {Barden}, {Bell}, {Bournaud}, {Brown}, {Caputi}, {Cassata}, {Challis}, {Chary}, {Cheung}, {Cirasuolo}, {Conselice}, {Roshan Cooray}, {Croton}, {Daddi}, {Dav{\'e}}, {de Mello}, {de Ravel}, {Dekel}, {Donley}, {Dunlop}, {Dutton}, {Elbaz}, {Fazio}, {Filippenko}, {Finkelstein}, {Frazer}, {Gardner}, {Garnavich}, {Gawiser}, {Gruetzbauch}, {Hartley}, {H{\"a}ussler}, {Herrington}, {Hopkins}, {Huang}, {Jha}, {Johnson}, {Kartaltepe}, {Khostovan}, {Kirshner}, {Lani}, {Lee}, {Li}, {Madau}, {McCarthy}, {McIntosh}, {McLure}, {McPartland}, {Mobasher}, {Moreira}, {Mortlock}, {Moustakas}, {Mozena}, {Nandra}, {Newman}, {Nielsen}, {Niemi}, {Noeske}, {Papovich},
  {Pentericci}, {Pope}, {Primack}, {Ravindranath}, {Reddy}, {Renzini}, {Rix}, {Robaina}, {Rosario}, {Rosati}, {Salimbeni}, {Scarlata}, {Siana}, {Simard}, {Smidt}, {Snyder}, {Somerville}, {Spinrad}, {Straughn}, {Telford}, {Teplitz}, {Trump}, {Vargas}, {Villforth}, {Wagner}, {Wandro}, {Wechsler}, {Weiner}, {Wiklind}, {Wild}, {Wilson}, {Wuyts}, \& {Yun}}]{Koekemoer2011}
{Koekemoer}, A.~M., {Faber}, S.~M., {Ferguson}, H.~C., {et~al.} 2011, \apjs, 197, 36, \dodoi{10.1088/0067-0049/197/2/36}

\bibitem[{{Koekemoer} {et~al.}(2013){Koekemoer}, {Ellis}, {McLure}, {Dunlop}, {Robertson}, {Ono}, {Schenker}, {Ouchi}, {Bowler}, {Rogers}, {Curtis-Lake}, {Schneider}, {Charlot}, {Stark}, {Furlanetto}, {Cirasuolo}, {Wild}, \& {Targett}}]{Koekemoer2013}
{Koekemoer}, A.~M., {Ellis}, R.~S., {McLure}, R.~J., {et~al.} 2013, \apjs, 209, 3, \dodoi{10.1088/0067-0049/209/1/3}

\bibitem[{{Kroupa}(2001)}]{Kroupa2001}
{Kroupa}, P. 2001, \mnras, 322, 231, \dodoi{10.1046/j.1365-8711.2001.04022.x}

\bibitem[{{Li} {et~al.}(2023){Li}, {Liu}, {Zhang}, {Tian}, {Fu}, {Li}, \& {Yan}}]{Li2023}
{Li}, J., {Liu}, C., {Zhang}, Z.-Y., {et~al.} 2023, \nat, 613, 460, \dodoi{10.1038/s41586-022-05488-1}

\bibitem[{{Li} {et~al.}(2025){Li}, {Diego}, {Meena}, {Lim}, {Fung}, {Levitskiy}, {Nianias}, {Palencia}, {Williams}, {Zhang}, {Amruth}, {Broadhurst}, {Chen}, {Filippenko}, {Kelly}, {Koekemoer}, {Perera}, {Sun}, {Williams}, {Windhorst}, {Yan}, \& {Zitrin}}]{Li2025}
{Li}, S.~K., {Diego}, J.~M., {Meena}, A.~K., {et~al.} 2025, arXiv e-prints, arXiv:2504.06992, \dodoi{10.48550/arXiv.2504.06992}

\bibitem[{{Lotz} {et~al.}(2017){Lotz}, {Koekemoer}, {Coe}, {Grogin}, {Capak}, {Mack}, {Anderson}, {Avila}, {Barker}, {Borncamp}, {Brammer}, {Durbin}, {Gunning}, {Hilbert}, {Jenkner}, {Khandrika}, {Levay}, {Lucas}, {MacKenty}, {Ogaz}, {Porterfield}, {Reid}, {Robberto}, {Royle}, {Smith}, {Storrie-Lombardi}, {Sunnquist}, {Surace}, {Taylor}, {Williams}, {Bullock}, {Dickinson}, {Finkelstein}, {Natarajan}, {Richard}, {Robertson}, {Tumlinson}, {Zitrin}, {Flanagan}, {Sembach}, {Soifer}, \& {Mountain}}]{Lotz2017}
{Lotz}, J.~M., {Koekemoer}, A., {Coe}, D., {et~al.} 2017, \apj, 837, 97, \dodoi{10.3847/1538-4357/837/1/97}

\bibitem[{{Mann} \& {Ebeling}(2012)}]{Mann2012}
{Mann}, A.~W., \& {Ebeling}, H. 2012, \mnras, 420, 2120, \dodoi{10.1111/j.1365-2966.2011.20170.x}

\bibitem[{{Massey} {et~al.}(1995){Massey}, {Lang}, {Degioia-Eastwood}, \& {Garmany}}]{Massey1995}
{Massey}, P., {Lang}, C.~C., {Degioia-Eastwood}, K., \& {Garmany}, C.~D. 1995, \apj, 438, 188, \dodoi{10.1086/175064}

\bibitem[{{Miralda-Escude}(1991)}]{Miralda-Escude1991}
{Miralda-Escude}, J. 1991, \apj, 379, 94, \dodoi{10.1086/170486}

\bibitem[{{M{\"u}ller} \& {Miralda-Escud{\'e}}(2025)}]{Muller2025}
{M{\"u}ller}, C.~V., \& {Miralda-Escud{\'e}}, J. 2025, \mnras, 536, 1579, \dodoi{10.1093/mnras/stae2652}

\bibitem[{{Oguri} {et~al.}(2018){Oguri}, {Diego}, {Kaiser}, {Kelly}, \& {Broadhurst}}]{Oguri2018}
{Oguri}, M., {Diego}, J.~M., {Kaiser}, N., {Kelly}, P.~L., \& {Broadhurst}, T. 2018, \prd, 97, 023518, \dodoi{10.1103/PhysRevD.97.023518}

\bibitem[{{Palencia} {et~al.}(2024){Palencia}, {Diego}, {Kavanagh}, \& {Mart{\'\i}nez-Arrizabalaga}}]{Palencia2024}
{Palencia}, J.~M., {Diego}, J.~M., {Kavanagh}, B.~J., \& {Mart{\'\i}nez-Arrizabalaga}, J. 2024, \aap, 687, A81, \dodoi{10.1051/0004-6361/202347492}

\bibitem[{{Paxton} {et~al.}(2011){Paxton}, {Bildsten}, {Dotter}, {Herwig}, {Lesaffre}, \& {Timmes}}]{Paxton2011}
{Paxton}, B., {Bildsten}, L., {Dotter}, A., {et~al.} 2011, \apjs, 192, 3, \dodoi{10.1088/0067-0049/192/1/3}

\bibitem[{{Paxton} {et~al.}(2013){Paxton}, {Cantiello}, {Arras}, {Bildsten}, {Brown}, {Dotter}, {Mankovich}, {Montgomery}, {Stello}, {Timmes}, \& {Townsend}}]{Paxton2013}
{Paxton}, B., {Cantiello}, M., {Arras}, P., {et~al.} 2013, \apjs, 208, 4, \dodoi{10.1088/0067-0049/208/1/4}

\bibitem[{{Paxton} {et~al.}(2015){Paxton}, {Marchant}, {Schwab}, {Bauer}, {Bildsten}, {Cantiello}, {Dessart}, {Farmer}, {Hu}, {Langer}, {Townsend}, {Townsley}, \& {Timmes}}]{Paxton2015}
{Paxton}, B., {Marchant}, P., {Schwab}, J., {et~al.} 2015, \apjs, 220, 15, \dodoi{10.1088/0067-0049/220/1/15}

\bibitem[{{Rihtar{\v{s}}i{\v{c}}} {et~al.}(2024){Rihtar{\v{s}}i{\v{c}}}, {Brada{\v{c}}}, {Desprez}, {Harshan}, {Noirot}, {Estrada-Carpenter}, {Martis}, {Abraham}, {Asada}, {Brammer}, {Iyer}, {Matharu}, {Mowla}, {Muzzin}, {Sarrouh}, {Sawicki}, {Strait}, {Willott}, {Gledhill}, {Markov}, \& {Tripodi}}]{CANUCS2024}
{Rihtar{\v{s}}i{\v{c}}}, G., {Brada{\v{c}}}, M., {Desprez}, G., {et~al.} 2024, arXiv e-prints, arXiv:2406.10332, \dodoi{10.48550/arXiv.2406.10332}

\bibitem[{{Salpeter}(1955)}]{Salpeter1955}
{Salpeter}, E.~E. 1955, \apj, 121, 161, \dodoi{10.1086/145971}

\bibitem[{{Schneider} {et~al.}(1992){Schneider}, {Ehlers}, \& {Falco}}]{Schneider1992}
{Schneider}, P., {Ehlers}, J., \& {Falco}, E.~E. 1992, {Gravitational Lenses}, \dodoi{10.1007/978-3-662-03758-4}

\bibitem[{{S{\'e}rsic}(1963)}]{Sersic1963}
{S{\'e}rsic}, J.~L. 1963, Boletin de la Asociacion Argentina de Astronomia La Plata Argentina, 6, 41

\bibitem[{{Sirianni} {et~al.}(2002){Sirianni}, {Nota}, {De Marchi}, {Leitherer}, \& {Clampin}}]{Sirianni2002}
{Sirianni}, M., {Nota}, A., {De Marchi}, G., {Leitherer}, C., \& {Clampin}, M. 2002, \apj, 579, 275, \dodoi{10.1086/342723}

\bibitem[{{Sirianni} {et~al.}(2000){Sirianni}, {Nota}, {Leitherer}, {De Marchi}, \& {Clampin}}]{Sirianni2000}
{Sirianni}, M., {Nota}, A., {Leitherer}, C., {De Marchi}, G., \& {Clampin}, M. 2000, \apj, 533, 203, \dodoi{10.1086/308628}

\bibitem[{{Steinhardt} {et~al.}(2020){Steinhardt}, {Jauzac}, {Acebron}, {Atek}, {Capak}, {Davidzon}, {Eckert}, {Harvey}, {Koekemoer}, {Lagos}, {Mahler}, {Montes}, {Niemiec}, {Nonino}, {Oesch}, {Richard}, {Rodney}, {Schaller}, {Sharon}, {Strolger}, {Allingham}, {Amara}, {Bah{\'e}}, {B{\oe}hm}, {Bose}, {Bouwens}, {Bradley}, {Brammer}, {Broadhurst}, {Ca{\~n}as}, {Cen}, {Cl{\'e}ment}, {Clowe}, {Coe}, {Connor}, {Darvish}, {Diego}, {Ebeling}, {Edge}, {Egami}, {Ettori}, {Faisst}, {Frye}, {Furtak}, {G{\'o}mez-Guijarro}, {Remolina Gonz{\'a}lez}, {Gonzalez}, {Graur}, {Gruen}, {Harvey}, {Hensley}, {Hovis-Afflerbach}, {Jablonka}, {Jha}, {Jullo}, {Kneib}, {Kokorev}, {Lagattuta}, {Limousin}, {von der Linden}, {Linzer}, {Lopez}, {Magdis}, {Massey}, {Masters}, {Maturi}, {McCully}, {McGee}, {Meneghetti}, {Mobasher}, {Moustakas}, {Murphy}, {Natarajan}, {Neyrinck}, {O'Connor}, {Oguri}, {Pagul}, {Rhodes}, {Rich}, {Robertson}, {Sereno}, {Shan}, {Smith}, {Sneppen}, {Squires}, {Tam}, {Tchernin}, {Toft}, {Umetsu}, {Weaver}, {van
  Weeren}, {Williams}, {Wilson}, {Yan}, \& {Zitrin}}]{Steinhardt2020}
{Steinhardt}, C.~L., {Jauzac}, M., {Acebron}, A., {et~al.} 2020, \apjs, 247, 64, \dodoi{10.3847/1538-4365/ab75ed}

\bibitem[{{Venumadhav} {et~al.}(2017){Venumadhav}, {Dai}, \& {Miralda-Escud{\'e}}}]{Venumadhav2017}
{Venumadhav}, T., {Dai}, L., \& {Miralda-Escud{\'e}}, J. 2017, \apj, 850, 49, \dodoi{10.3847/1538-4357/aa9575}

\bibitem[{{Weisenbach} {et~al.}(2024){Weisenbach}, {Anguita}, {Miralda-Escud{\'e}}, {Oguri}, {Saha}, \& {Schechter}}]{Weisenbach2024}
{Weisenbach}, L., {Anguita}, T., {Miralda-Escud{\'e}}, J., {et~al.} 2024, arXiv e-prints, arXiv:2404.08094, \dodoi{10.48550/arXiv.2404.08094}

\bibitem[{{Weisz} {et~al.}(2015){Weisz}, {Johnson}, {Foreman-Mackey}, {Dolphin}, {Beerman}, {Williams}, {Dalcanton}, {Rix}, {Hogg}, {Fouesneau}, {Johnson}, {Bell}, {Boyer}, {Gouliermis}, {Guhathakurta}, {Kalirai}, {Lewis}, {Seth}, \& {Skillman}}]{Weisz2015}
{Weisz}, D.~R., {Johnson}, L.~C., {Foreman-Mackey}, D., {et~al.} 2015, \apj, 806, 198, \dodoi{10.1088/0004-637X/806/2/198}

\bibitem[{{Welch} {et~al.}(2022{\natexlab{a}}){Welch}, {Coe}, {Diego}, {Zitrin}, {Zackrisson}, {Dimauro}, {Jim{\'e}nez-Teja}, {Kelly}, {Mahler}, {Oguri}, {Timmes}, {Windhorst}, {Florian}, {de Mink}, {Avila}, {Anderson}, {Bradley}, {Sharon}, {Vikaeus}, {McCandliss}, {Brada{\v{c}}}, {Rigby}, {Frye}, {Toft}, {Strait}, {Trenti}, {Sharma}, {Andrade-Santos}, \& {Broadhurst}}]{Welch2022a}
{Welch}, B., {Coe}, D., {Diego}, J.~M., {et~al.} 2022{\natexlab{a}}, \nat, 603, 815

\bibitem[{{Welch} {et~al.}(2022{\natexlab{b}}){Welch}, {Coe}, {Zackrisson}, {de Mink}, {Ravindranath}, {Anderson}, {Brammer}, {Bradley}, {Yoon}, {Kelly}, {Diego}, {Windhorst}, {Zitrin}, {Dimauro}, {Jim{\'e}nez-Teja}, {Abdurro'uf}, {Nonino}, {Acebron}, {Andrade-Santos}, {Avila}, {Bayliss}, {Ben{\'\i}tez}, {Broadhurst}, {Bhatawdekar}, {Brada{\v{c}}}, {Caminha}, {Chen}, {Eldridge}, {Farag}, {Florian}, {Frye}, {Fujimoto}, {Gomez}, {Henry}, {Hsiao}, {Hutchison}, {James}, {Joyce}, {Jung}, {Khullar}, {Larson}, {Mahler}, {Mandelker}, {McCandliss}, {Morishita}, {Newshore}, {Norman}, {O'Connor}, {Oesch}, {Oguri}, {Ouchi}, {Postman}, {Rigby}, {Ryan}, {Sharma}, {Sharon}, {Strait}, {Strolger}, {Timmes}, {Toft}, {Trenti}, {Vanzella}, \& {Vikaeus}}]{Welch2022b}
{Welch}, B., {Coe}, D., {Zackrisson}, E., {et~al.} 2022{\natexlab{b}}, \apjl, 940, L1.
\newblock \doarXiv{2208.09007}

\bibitem[{Williams(in prep.)}]{Williams+inprep}
Williams, H. in prep.

\bibitem[{{Williams} {et~al.}(2024){Williams}, {Kelly}, {Treu}, {Amruth}, {Diego}, {Li}, {Meena}, {Zitrin}, {Broadhurst}, \& {Filippenko}}]{Williams2024}
{Williams}, L. L.~R., {Kelly}, P.~L., {Treu}, T., {et~al.} 2024, \apj, 961, 200, \dodoi{10.3847/1538-4357/ad1660}

\bibitem[{{Windhorst} {et~al.}(2023){Windhorst}, {Cohen}, {Jansen}, {Summers}, {Tompkins}, {Conselice}, {Driver}, {Yan}, {Coe}, {Frye}, {Grogin}, {Koekemoer}, {Marshall}, {O'Brien}, {Pirzkal}, {Robotham}, {Ryan}, {Willmer}, {Carleton}, {Diego}, {Keel}, {Porto}, {Redshaw}, {Scheller}, {Wilkins}, {Willner}, {Zitrin}, {Adams}, {Austin}, {Arendt}, {Beacom}, {Bhatawdekar}, {Bradley}, {Broadhurst}, {Cheng}, {Civano}, {Dai}, {Dole}, {D'Silva}, {Duncan}, {Fazio}, {Ferrami}, {Ferreira}, {Finkelstein}, {Furtak}, {Gim}, {Griffiths}, {Hammel}, {Harrington}, {Hathi}, {Holwerda}, {Honor}, {Huang}, {Hyun}, {Im}, {Joshi}, {Kamieneski}, {Kelly}, {Larson}, {Li}, {Lim}, {Ma}, {Maksym}, {Manzoni}, {Meena}, {Milam}, {Nonino}, {Pascale}, {Petric}, {Pierel}, {Polletta}, {R{\"o}ttgering}, {Rutkowski}, {Smail}, {Straughn}, {Strolger}, {Swirbul}, {Trussler}, {Wang}, {Welch}, {B. Wyithe}, {Yun}, {Zackrisson}, {Zhang}, \& {Zhao}}]{Windhorst2023}
{Windhorst}, R.~A., {Cohen}, S.~H., {Jansen}, R.~A., {et~al.} 2023, \aj, 165, 13, \dodoi{10.3847/1538-3881/aca163}

\bibitem[{{Yan} {et~al.}(2023){Yan}, {Ma}, {Sun}, {Wang}, {Kelly}, {Diego}, {Cohen}, {Windhorst}, {Jansen}, {Grogin}, {Beacom}, {Conselice}, {Driver}, {Frye}, {Coe}, {Marshall}, {Koekemoer}, {Willmer}, {Robotham}, {D'Silva}, {Summers}, {Nonino}, {Pirzkal}, {Ryan}, {Ortiz}, {Tompkins}, {Bhatawdekar}, {Cheng}, {Zitrin}, \& {Willner}}]{Yan2023}
{Yan}, H., {Ma}, Z., {Sun}, B., {et~al.} 2023, \apjs, 269, 43, \dodoi{10.3847/1538-4365/ad0298}

\bibitem[{{Zackrisson} {et~al.}(2024){Zackrisson}, {Hultquist}, {Kordt}, {Diego}, {Nabizadeh}, {Vikaeus}, {Meena}, {Zitrin}, {Volpato}, {Lundqvist}, {Welch}, {Costa}, \& {Windhorst}}]{Zackrisson2024}
{Zackrisson}, E., {Hultquist}, A., {Kordt}, A., {et~al.} 2024, \mnras, \dodoi{10.1093/mnras/stae1881}

\end{thebibliography}
\bibliographystyle{aasjournal}

\end{document}